%% file: main_clean.tex
\documentclass[a4paper,11pt]{elsarticle}
\usepackage[margin=1in]{geometry}
\usepackage{lineno}
\usepackage{paralist}
\usepackage{amsmath,amssymb,amsfonts}
\usepackage{verbatim}
\usepackage{graphicx}
\usepackage{rotating}
\usepackage{float}
\usepackage{xcolor, soul}
\usepackage{hyperref}
\usepackage{subfigure}
\usepackage[normalem]{ulem}
\usepackage[bottom]{footmisc}
\usepackage{adjustbox}
\usepackage{graphics}
\title{Impact of Simultaneous Stellar Modeling Uncertainties on the Tip of the Red Giant Branch for Axion-Election Coupling}

\DeclareRobustCommand{\okina}{%
  \raisebox{\dimexpr\fontcharht\font`A-\height}{%
    \scalebox{0.8}{`}%
  }%
}
\pdfoutput=1
\allowdisplaybreaks

\newcommand{\msun}{{\rm M}_\odot}

\author[1]{Mitchell T.~Dennis}
\affiliation[1]{organization={Institute for Astronomy, University of Hawaii at Manoa},
            addressline={2680 Woodlawn Drive}, 
            city={Honolulu},
            postcode={96822}, 
            state={HI},
            country={USA}}

\author[2]{and Jeremy Sakstein}
\affiliation[2]{organization={Department of Physics \& Astronomy, University of Hawaii at Manoa},
            addressline={Watanabe Hall 2505 Correa Road}, 
            city={Honolulu},
            postcode={96822}, 
            state={HI},
            country={USA}}

\date{\today}
\begin{document}

\begin{abstract}
We present a novel method for incorporating the effects of stellar modeling uncertainties into constraints on the axion-electron coupling constant found using the observed calibration of the tip of the red giant branch (TRGB) I band magnitude $M_I$.~We simulate grids of models with varying initial stellar mass, helium abundance, metallicity, and axion-electron coupling $\alpha_{26}= 10^{26} g^2_{ae}/4\pi$ but different (fixed) mixing lengths and mass loss efficiencies.~We then train separate machine learning emulators to predict $M_I$ as a function of the varying parameters for each grid.~Our emulators enable the use of Markov Chain Monte Carlo simulations where $\alpha_{26}$ is varied simultaneously with the stellar parameters.~One of our grids yields a bound $\alpha_{26}\leq 0.75$ at the 95\% confidence limit,  a factor of $\sim3.7$ weaker than previous bounds;~while the other grid yields $\alpha_{26}\leq1.58$ at the 95\% confidence limit, a factor $\sim7.8$ weaker than previous bounds.~We demonstrate that the different values we find are due to covariances between stellar and axion physics that are not accounted for by single parameter variations.~Our results suggest that the bound on $\alpha_{26}$ derived using empirical calibrations of the TRGB I band magnitude need to be reevaluated using simultaneous parameter variation.~Alternative methods that use the bolometric luminosity instead of $M_I$ are more robust because they are not reliant upon theoretical predictions of the effective temperature.
\end{abstract}

\include{newcommands}

\maketitle
\section{Introduction}
\label{sec:intro}
The extreme environments inside stellar objects are impossible to replicate on Earth, making stars unique laboratories for testing theories of physics beyond the standard model \cite{1996slfp.book.....R}, particularly those with weak couplings and light masses ($<\!10$keV).~Indeed, stars have been used to search for dark matter (DM) candidates such as hidden photons \cite{Ayala:2015juy, 2020JCAP...11..029A}, WIMPS \cite{2021A&A...651A.101L}, QCD axions and axion-like particles \cite{Capozzi2020, Straniero2020}, and new interactions that arise in theories beyond the standard model of particle physics such as those where the neutrino has a large magnetic dipole moment \cite{Viaux2013, 2013PhRvL.111w1301V, 2015APh....70....1A}.~Previous works have not been able to fully account for the uncertainties and degeneracies due to stellar input physics.~Accounting for these requires statistical methods such as Markov Chain Monte Carlo (MCMC) analyses that vary stellar and new physics parameters simultaneously.~MCMC algorithms sample the parameter space to find the region of maximum likelihood, and converge to it in a reasonable timescale provided that the evaluation time per sample is of order a second or shorter.~Unfortunately, the long run times of stellar modeling software (of order hours) have prohibited its use.~In this work, we overcome this challenge by utilizing machine learning (ML) as an emulator to reduce the run time of stellar modeling software to milliseconds, enabling the use of MCMC.~We focus on light axion-like particles --- hereafter referred to as \textit{axions} --- ($m_a<10$keV) as an application of this method to reevaluate the  constraints on the axion-electron coupling coming from observations of the tip of the red giant branch (TRGB) stars\footnote{Axion-like particles also couple to photons, but the effects of this coupling are negligible in TRGB stars \cite{Altherr:1993zd} so we do not consider it in this work.}.~We chose this as a case study because TRGB stars provide the strongest constraints on this coupling (at light masses).~

Light axions can be produced in large quantities in the high temperature and density environments of stellar cores  through semi-Compton scattering and bremsstrahlung processes \cite{1995PhRvD..51.1495R}.~The axions subsequently free-stream out of the star, providing a novel source of energy loss analogous to neutrinos, acting as an additional cooling mechanism for the stellar core.~This increases the requisite mass required for the core to reach the $10^8$K which triggers the helium flash, resulting in an increase in the brightness of  the TRGB, and consequentially a decrease in the I band magnitude $M_I$ which is the empirical observable~\cite{Serenelli2017, Capozzi2020}.~The TRGB $M_I$ has been calibrated in many systems, partly because it is a standard candle \cite{1999IAUS..183...48S, 2019ApJ...882...34F, 2021ApJ...908L...6R}.~The axion-electron coupling can be constrained by directly comparing the calibrations from observations with theoretical predictions from stellar structure codes.~The theoretical $M_I$ is subject to uncertainties from stellar input physics \cite{2010ApJ...719..865S,2022MNRAS.514.3058S} and empirical bolometric corrections needed to convert  the outputs of stellar structure codes to $M_I$.

One of the major sources of theoretical uncertainty in $M_I$ is the effective temperature $T_{\rm eff}$, which is required for calculating the bolometric corrections.~An alternate method \cite{Straniero2020, Troitsky:2024keu} that compares the observed bolometric luminosity $M_{\rm bol}$ with observations bypasses this uncertainty so has the advantage that it is more robust to theoretical uncertainties that affect $T_{\rm eff}$. In addition, multiple systems can be used to create a cumulative likelihood, enhancing the statistical power.


In this work, we will adopt the method where $M_I$ is used as the observational probe, so we cannot draw conclusions about the $M_{\rm bol}$ method.~In the $M_I$ method, theoretical errors in the axion bounds are incorporated by first fixing a fiducial model that reproduces the observed properties of the system under consideration and then varying each input parameter individually over the range implied by other measurements of the system to find the variation in $M_I$.~The uncertainty in $M_I$ is then calculated by assuming that each parameter has a top-hat probability distribution, and convolving these to find an approximate Gaussian distribution for $M_I$ whose standard deviation is taken to be the uncertainty.The most recent application of this method is by reference \cite{Capozzi2020}, who applied it to the LMC (Y19) to obtain the bound $\alpha_{26}<0.124$ (95\% C.L.), which is the strongest they find among the systems they studied.(The $M_{\rm bol}$ method yields a stronger bound $\alpha_{26}< 0.02$ \cite{Troitsky:2024keu}.)~Individual parameter variation has the drawback that it does not account for degeneracies, which only manifest when all parameters  --- including $\alpha_{26}$ --- are varied simultaneously.~Failing to account for degeneracies may lead to overestimation of the final bounds.~The novelty of our method lies in overcoming this limitation by enabling a simultaneous variation of parameters using MCMC.

Our method for incorporating these errors into MCMC is as follows.~First, we run a grid of stellar models with varying stellar input parameters and axion-electron coupling $\alpha_{26}=10^{26}g_{ae}^2/4\pi$ (see \ref{sec:axions} for the definition of this).~We then train a ML emulator on this grid to predict the (color-corrected) TRGB $M_I$ and the error due to the bolometric corrections as a function of the parameters.~This is used to generate theoretical predictions in an MCMC code that compares them with empirical calibrations. 

Due to limited computing time, we were only able to vary the mass, metallicity, helium mass fraction, and axion-electron coupling, so we ran two grids with different fixed input physics.~The input physics in the first is similar to that used by \cite{Viaux2013,Capozzi2020} while the second adopts different values for the mixing length and mass loss efficiency within the range that \cite{Viaux2013,Capozzi2020} varied.~This enables us to explore more than one location in parameter space to see if our constraint holds across different values $\alpha_{\rm MLT}$ and mass loss parameter space without running a large number of models. This allows us to assess if these parameters do not have substantial covariance across parameter space as previous claimed.~We applied our method to both grids in separate analyses to asses the previous bounds derived using single parameter variation.~Our analyses mirror those of \cite{Viaux2013,Capozzi2020}.~We impose Gaussian priors on the parameters that we vary corresponding to the range varied by \cite{Viaux2013,Capozzi2020}, but we allow the parameters to vary simultaneously via MCMC.~Our analysis of the first grid yields a bound $\alpha_{26}<0.75$ (95\% C.L.), a factor of $3.7$ weaker than  \cite{Viaux2013,Capozzi2020} while our analysis of the second grid yields $\alpha_{26}<1.58$ (95\% C.L.), a factor of $7.8$ weaker than previous studies.~We investigate the cause of this discrepancy by studying the tip variation $\Delta M_{I, tip}$ as a function of mixing length, mass loss, and axion-electron coupling, finding a strong correlation that is not accounted for by the single parameter variation. $\Delta M_{I, tip}$ is defined as the difference in the maximum $M_I$ between the input physics for benchmarks 1 and 2 for matching M, Y, Z, and $\alpha_{26}$.

\begin{table*}
\centering
\resizebox{\textwidth}{!}{
\begin{tabular}{|c|c|c|c|c|}
    \hline
    Target & Reference & $M_I$ & $\alpha_{26}$ Benchmark 1 & $\alpha_{26}$ Benchmark 2 \\
    \hline
    LMC & Y19 \cite{2019ApJ...886...61Y} & $-3.958 \pm 0.046$ & 0.57 & 1.55 \\
    \hline
    \dots & F20 \cite{2020ApJ...891...57F} & $-4.047 \pm 0.045$ & 0.75 & 1.58 \\
    \hline
    NGC 4258 & \cite{2017ApJ...835...28J} Updated by \cite{Capozzi2020}  & $-4.027 \pm 0.055$ & 0.71 & 1.56 \\
    \hline
    $\omega$-Centauri & \cite{2001ApJ...556..635B} Updated by  \cite{Capozzi2020} & $-3.96 \pm 0.05$ & 0.59 & 1.55 \\
    \hline
    \dots & \cite{2001ApJ...556..635B} Updated by this work & $-4.037 \pm 0.045$ & 0.74 & 1.57 \\
    \hline
\end{tabular}}
\caption{The empirical $M_I$ calibrations, their errors, and the axion bounds obtained for benchmarks 1 and 2, respectively.~The values for NGC 4258 and $\omega$-Centauri were updated by \cite{Capozzi2020} and these updated values are the ones given in the table. These values are all shifted to their reference color $V-I$=1.8. We give two values for $\omega$-Centauri, the one used by \cite{Capozzi2020} (which used a distance of $5.24 \pm 0.05$ kpc from \cite{Baumgardt2019}) and one updated by this work using the same procedure as \cite{Capozzi2020}, but instead with the updated distance $5.426 \pm 0.047$ from \cite{Baumgardt2021}.}
\label{tbl:calibrations}
\end{table*}

This paper is organized as follows.~In section \ref{sec:theory} we describe the stellar structure code, input physics, and grid of models used to train the ML emulators.~In section \ref{sec:ML} we describe the ML methods we use to train our emulators.~In section \ref{sec:MCMC} we use MCMC to constrain the axion-electron couplings using empirical TRGB $M_I$ calibrations in the Large Magellanic Cloud (LMC), NGC 4258, and $\omega$-Centauri.~We discuss the implications of our results and conclude in section \ref{sec:conc}.~In \ref{sec:axions} we briefly describe axions coupled to electrons for the unfamiliar reader and present our implementation of their energy loss rate into our stellar structure code.~A reproduction package accompanies this work and can be found at the following URL:~\url{https://zenodo.org/record/7896061}.~This includes our modifications to the stellar structure code, our entire grid of models and the stellar structure code inputs (inlists) needed to reproduce them, our ML emulators, and our MCMC scripts used to produce the results presented here.

\section{Theoretical Tip of the Red Giant Branch Models}
\label{sec:theory}

\subsection{Stellar Structure Code and Input Physics}
\label{sec:input_physics}
Our simulations were performed using the stellar structure code Modules for Experiments in Stellar Astrophysics (MESA) version {12778} \cite{Paxton2011,Paxton2013,Paxton2015,Paxton2018,Paxton2019,2022arXiv220803651J} to account for axion losses by modifying the neutrino loss rate i.e.,~we treated axions as an additional form of neutrino loss.~For a given set of parameters, each model was evolved from the pre-main-sequence to the onset of the Helium flash (defined as the point where the power from helium burning exceeds $10^6$ ergs/s).~As mentioned above, we were unable to vary all input physics due to limited computing time, so we fixed some quantities to fiducial values.~These match those used by \cite{Capozzi2020} and are as follows. \\

\noindent\textbf{Conduction Opacity:}~We use the most recent relations of Cassisi et al.~(2007) \cite{Cassisi2007}, which completely covers the range of thermal conditions expected to be relevant for degenerate electrons in the cores of low-mass, metal-poor stars;~is appropriate for arbitrary chemical compositions;~and includes the contributions of electron-ion  and
electron-electron scattering, accounting for partial electron degeneracy.\newline
\textbf{Radiative Opacity}:~We use the OPAL type 2 radiative opacity tables \cite{1996ApJ...464..943I}.\newline
\textbf{Nuclear Reaction Rates:~}We use the JINA {REACLIB} tables \cite{2010ApJS..189..240C}.\newline
\textbf{Nuclear Screening Factors}:~We use the prescription of Chugunov et al.~(2007) \cite{Chugunov2007}, which provides a smooth parameterization for the intermediate screening regime and reduces to the weak and strong limits at small and large plasma parameters respectively.~The MESA implementation is appropriate for modeling arbitrary multi-component plasmas.\newline
\textbf{Equation of State:}~We use the OPAL 2005 equation of state \cite{Rogers2002}.\newline
\textbf{Neutrino Loss Rates}~We use the neutrino loss rates of Itoh et al.~(1996) \cite{Itoh1996}.\newline
\textbf{Initial Elemental Abundances:}~We used the initial elemental abundances reported by \cite{1998SSRv...85..161G} (GS98).\newline
\textbf{Mass Loss:}~Mass is lost on the red giant branch (RGB) according to the Reimers prescription ($\dot{M}\propto \eta RL/M$) \cite{1975MSRSL...8..369R}.~We used two different values for efficiency parameter $\eta$ given in table \ref{tab:benchmarks}.\newline
\textbf{Convective Mixing:}~We use Mixing Length theory according to the prescription of Cox  \cite{1968pss..book.....C} with two different values of the mixing parameter $\alpha_{\rm MLT}$ given in table \ref{tab:benchmarks}.

\subsection{Theoretical $M_I$ Calculation}
\label{sec:mi_correlation}
Using the stellar structure code discussed above, we explored the effects of varying stellar and axions parameters on the TRGB $M_I$. Theoretically, $M_I$ is calculated by applying empirically-calibrated bolometric corrections (BCs) to the output of stellar structure codes (e.g., \cite{Serenelli2017,Viaux2013}).~These BCs take $T_\textrm{eff}$, iron abundance [Fe/H], luminosity $L$, and surface gravity $\log_{10}(g)$ ($g=GM/r^2$) as inputs.~Thus, any stellar or new physics parameters that alter these values at the TRGB are a source of theoretical uncertainty.~In this work, we use the bolometric correction program provided by Worthey \& Lee (WL) \cite{WortheyLee2011}.\footnote{Because bolometric corrections can produce systematic errors, our reproduction package \cite{dennis_mitchell_t_2023_7896061} includes a ML emulator (not used in this work) that predicts the four values passed to the Worthey \& Lee bolometric corrections that can be used to calculate $M_I$ using other bolometric corrections e.g., MARCS \cite{2008A&A...486..951G} or PHOENIX \cite{2008ApJS..178...89D}.~We comment for users interested in this that training an emulator on these bolometric corrections applied to our grid (also included in the reproduction package) will  yield more accurate results than applying corrections to the results of this emulator.} This code computes the $V-I$ color and the bolometric correction from the V band (and their respective errors). From these values and our model luminosity, we calculate $M_I$ and its associated error ($\Delta M_I$).

To exemplify the effects of the stellar and axion uncertainties considered in this work, and correlations between them, we explored the effect of initial mass ($M$), initial metallicity ($Z$), mixing length ($\alpha_{\rm MLT}$), the Reimer's wind loss scaling factor ($\eta$), and the axion-electron coupling constant ($\alpha_{26}$) on the evolution of RGB stars and their corresponding TRGB $M_I$ values.~The results are shown in Fig.~\ref{fig:exploration}.~The interplay among these parameters is complex and non-linear --- especially after bolometric corrections are applied --- so in what follows we highlight only the leading qualitative effects.~These are: 
\\

\noindent \textbf{Mass:}The bolometric luminosity at the TRGB is largely set by the nearly constant Helium core mass at helium ignition, so it only varies weakly with $M$. However, changes in $M$ alter the surface gravity and $T_{\rm eff}$, which in turn modify the bolometric corrections and thus the predicted tip $M_I$. \newline
\textbf{Metallicity:}Increasing $Z$ raises the envelope opacity and cools the stellar surface, shifting the bolometric correction. As a result, $M_I$ has a non-linear dependence on $Z$.~The situation is further complicated by axion emission, whose rate carries an explicit dependence on $Z$ (Eqs.~\eqref{eq:ndBrem}--\eqref{eq:kappa}).\newline
\textbf{Helium Abundance:}As $Y$ increases, it creates a heavier helium core with a hotter central temperature, leading to a brighter helium flash.~Axion energy losses also depend on $Y$ via Eqns.~\ref{eq:ndBrem}--\ref{eq:kappa}.\newline
\textbf{Convective mixing:}A higher $\alpha_{\rm MLT}$ makes convection more efficient, reducing the superadiabatic gradient and raising $T_{\rm eff}$. Because bolometric corrections depend sensitively on $T_{\rm eff}$, this in turn alters the theoretical prediction for $M_I$ in a non-linear manner.\newline
\textbf{Wind loss:}Larger $\eta$ strips the hydrogen-rich envelope on the RGB, lowering the envelope mass and surface gravity. This typically leads to a fainter TRGB (higher $M_I$) via its impact on the bolometric correction.\newline
\textbf{Axion-electron coupling:}Axions provide an additional channel for energy loss, cooling the helium core. This delays the onset of the helium flash, allowing the inert core to grow more massive through continued hydrogen shell burning. A larger core ignites helium more violently, resulting in a brighter TRGB and thus a lower (more negative) $M_I$.~Two ancillary effects are that stronger axion losses reduce the star's radius and increase its effective temperature \cite{Frieman:1987ui}.\newline

These effects are exemplified in Figure ~\ref{fig:exploration}, where we show Hertzsprung-Russell (HR) tracks tracks and $M_I$ at the TRGB as a function of these parameters. The plots not only show how $M_I$ depends on these parameters but also demonstrate that the uncertainties are correlated. The final stages of the HR tracks are plotted on the left (terminating at the upright and inverted triangles), and $M_I$ magnitudes are plotted on the right as a function of $V-I$. Down the rows, the panels vary a combination of three of our input parameters. For each row where a particular parameter is not varied, that parameter is held fixed. The fixed values of the parameters are $M = 1.0$, $Y = 0.252$, $Z = 0.00012$, $\alpha_{\rm MLT} = 1.896$, and $\eta = 0.0$ (e.g. for row one $\alpha_{\rm MLT} = 1.896$ and $\eta = 0.0$), and all other physics are fixed to the values listed above in \S \ref{sec:input_physics}. In the left column, varying colors represent either a variance in $M$ or $Z$; the linestyles represent either $Z$, $\alpha_{\rm MLT}$, or $\eta$; and the purple and orange triangles (upright and inverted, respectively) represent $\alpha_{26}$. In the right column panels, varying colors also represent either a variance in $M_I$ or $Z$; different shapes represent either $Z$, $\alpha_{\rm MLT}$, or $\eta$; and different point sizes represent differences in $\alpha_{26}$. We  do not show plots with varying $Y$ as its affect on the TRGB is roughly linear as shown by \cite{dennis2025machinelearningtipred} in their Figure 1. We instead focus on more complex relationships and correlations.

To explain what we mean by covariance/correlation between parameters, we direct the reader to the second and fourth rows of the right hand column in Figure \ref{fig:exploration}. The primary difference between these two panels is that the second row varies $M$ and the fourth row varies $Z$. In the panels which vary $M$, there appears to be little difference in the $M_I$ between the parameters (with the exception of $\alpha_{26}$); the gaps between the individual points of different shapes are roughly constant; and the slope of the line between points of the same color (but different shapes) do not change. However this is no longer the case when $Z$ is varied, in the panels which vary $Z$, the gaps between points of the same color shift significantly and produce different slopes. We describe the most important  correlations below: \\

\noindent\textbf{Axion-electron coupling and Metallicity:} The panels where $\alpha_{26}$ and $Z$ vary simultaneously (rows 1, 4, and 5 of Fig.~1) show the largest spread in TRGB $M_I$. Metallicity increases the envelope opacity and enters explicitly into the axion emissivity (Eqs.~\eqref{eq:ndBrem}--\eqref{eq:kappa}), while a larger $\alpha_{26}$ delays helium ignition and increases the core mass.~Together these effects lead to a strong correlation between $Z$ and $\alpha_{26}$ in their impact on $M_I$. \newline
\textbf{Axion-electron coupling and Mass:} Stellar mass mainly affects surface properties such as $T_{\rm eff}$ and $\log g$, which enter the bolometric correction, while axion cooling (through $\alpha_{26}$) delays helium ignition and increases the core mass, raising the bolometric luminosity. When both parameters vary simultaneously, the resulting changes in bolometric corrections and luminosity combine to produce the spread in $M_I$ seen in Fig.~1.\newline
\textbf{Metallicity and Mixing Length:} In the fourth row of the right column of Fig.~1, the spread in $M_I$ (and $V-I$) is minimal at low $Z$, but grows substantially at higher $Z$. Higher metallicity increases the envelope opacity, so the efficiency of convection (set by $\alpha_{\rm MLT}$) becomes more important for energy transport. This in turn strongly affects $T_{\rm eff}$, the bolometric correction, and thus $M_I$.\newline
\textbf{Metallicity and Wind loss:} The combination of higher $Z$ and larger $\eta$ produces an enhanced spread in $M_I$ (see the bottom right panel). At higher metallicity the stellar envelope is more opaque, so stripping mass through winds alters the outer structure and $T_{\rm eff}$ more strongly, compounding the impact on the bolometric correction and $M_I$.\newline

These correlations motivate simultaneous parameter variation when accounting for uncertainties in data analyses.~Our work is a first step towards this.

\begin{figure*}[h]
    \centering
    \begin{subfigure}
        \centering
        \includegraphics[scale=0.50]{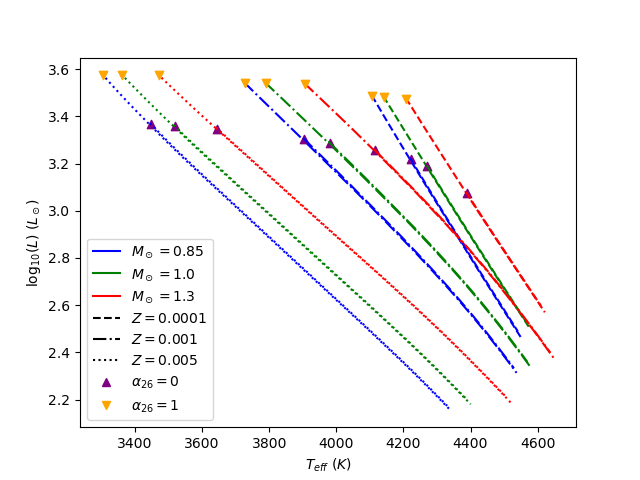}
        \includegraphics[scale=0.47]{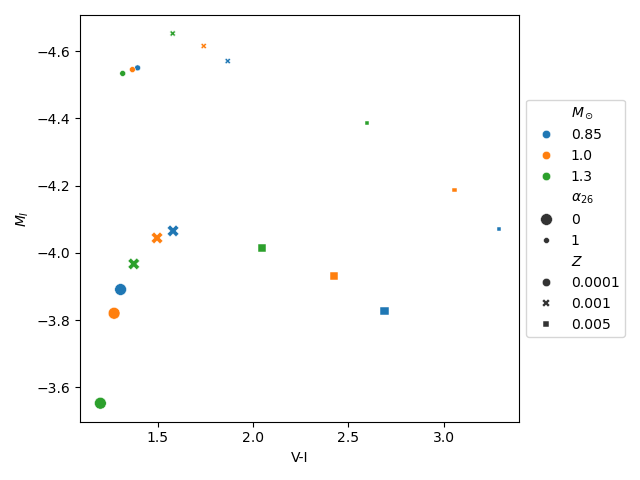}
    \end{subfigure}
    \begin{subfigure}
        \centering
        \includegraphics[scale=0.50]{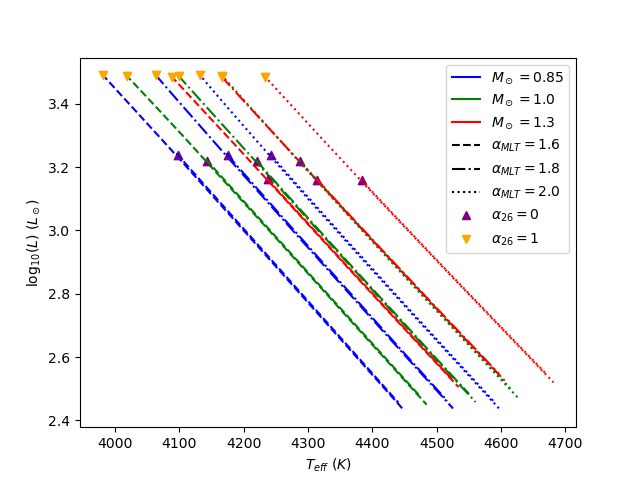}
        \includegraphics[scale=0.47]{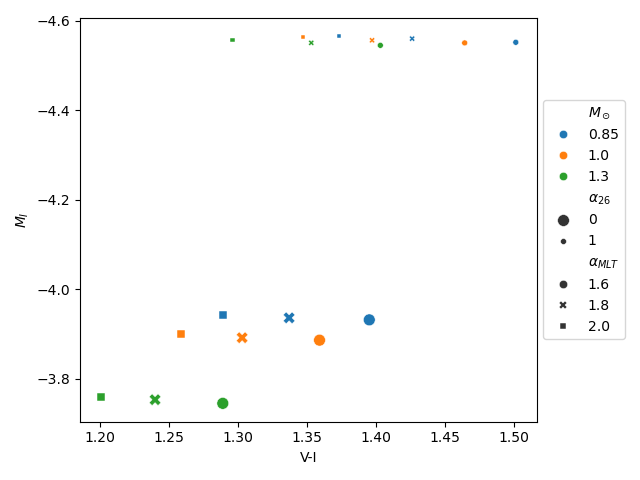}
    \end{subfigure}
    \begin{subfigure}
        \centering
        \includegraphics[scale=0.50]{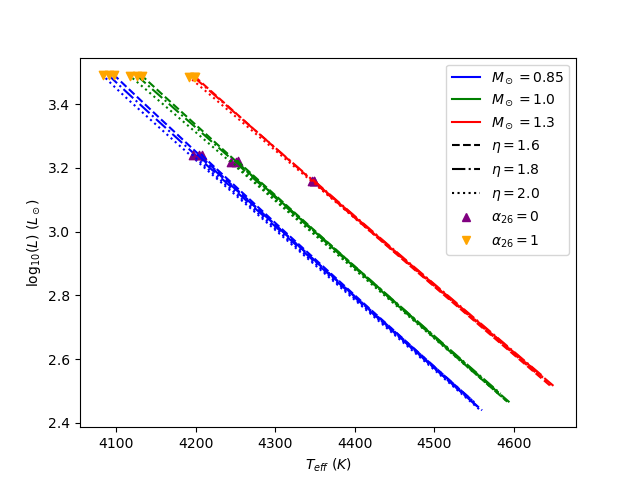}
        \includegraphics[scale=0.47]{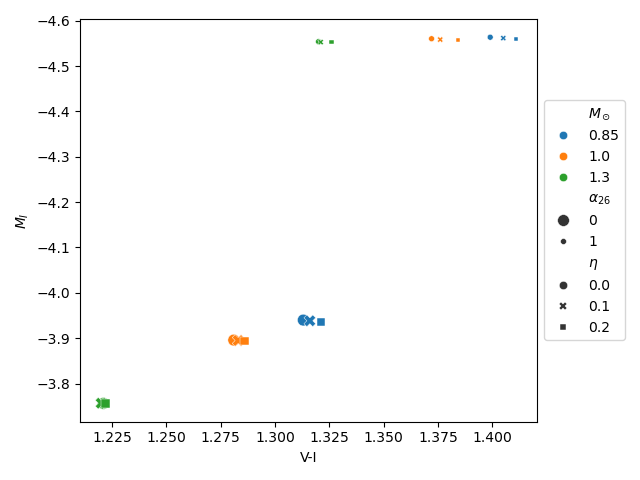}
    \end{subfigure}
\end{figure*}
\clearpage
\begin{figure*}[ht!]
    \begin{subfigure}
        \centering
        \includegraphics[scale=0.50]{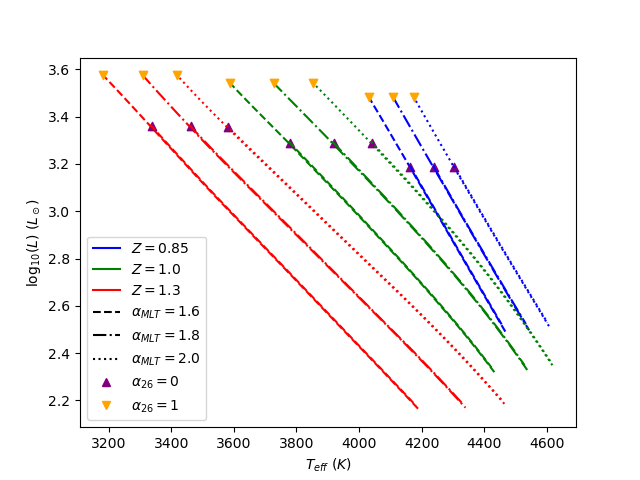}
        \includegraphics[scale=0.47]{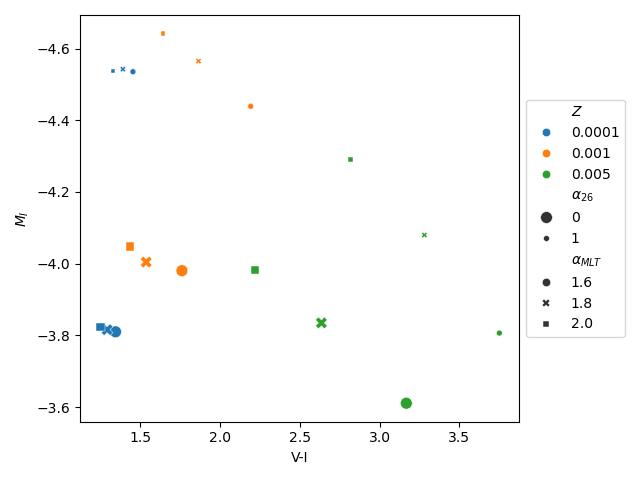}
    \end{subfigure}
    \begin{subfigure}
        \centering
        \includegraphics[scale=0.50]{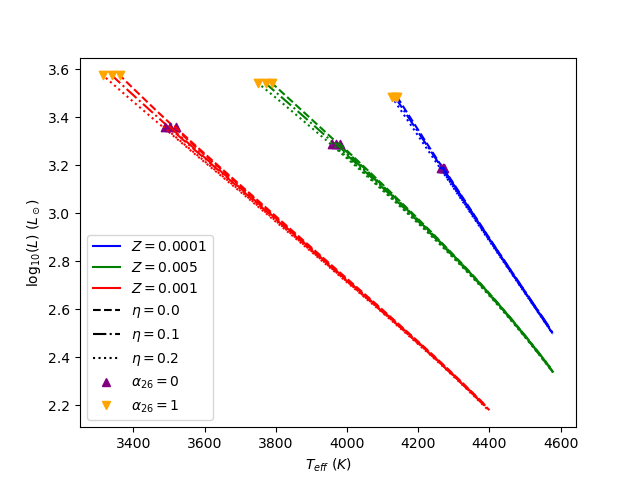}
        \includegraphics[scale=0.47]{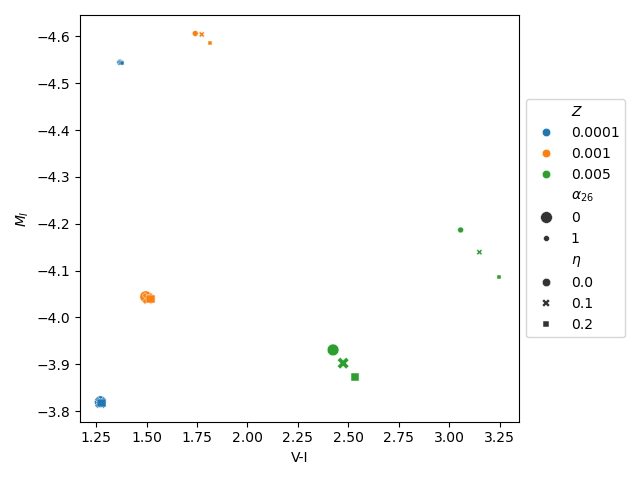}
    \end{subfigure}
    \caption{Plots showing the difference in the stellar evolution tracks in the HR diagram close to the TRGB (left column) and the $M_I$ values at the TRGB (right column) for different sets of varying parameters (rows). Further details are given in the text.} 
    \label{fig:exploration}
\end{figure*}

\subsection{Benchmark Models}

\begin{table}[ht]
    \centering
    \begin{tabular}{|c|c|c|}\hline
       Benchmark  & Wind Loss Efficiency $\eta$ & Mixing Length $\alpha_{\rm MLT}$ \\\hline
        1 & 0.0 & 1.892\\\hline
        2 & 0.1 & 1.800
        \\\hline 
    \end{tabular}
    \caption{Benchmark Models used in our data analyses.}
    \label{tab:benchmarks}
\end{table}

Due to computational limits, we could not vary all of the parameters analyzed above. Instead, to facilitate our study of the effect of covariances upon the axion-electron coupling bound we considered two benchmark models given in Table~\ref{tab:benchmarks}.~Benchmark 1 is similar to the benchmark model of \cite{Viaux2013}.~Benchmark 2 differs in that it includes wind loss and uses a smaller mixing length. When we performed a solar calibration for the mixing length for our MESA version using identical input physics as \cite{Viaux2013}, we found $\alpha_{\rm MLT} = 1.884$, within 0.5\% of 1.892, and between the two values used by \cite{Serenelli2017} (1.83 and 2.02). \cite{Serenelli2017} does not account for wind loss (i.e. $\eta = 0$), and reference \cite{Viaux2013} varied the mixing length and wind loss efficiency individually about Benchmark 1 to determine their associated uncertainties. We note that wind loss is not important until post main-sequence, and therefore does not affect the solar mixing length calibrations.~This single-parameter estimation accounts for variances but not covariances i.e., the uncertainties due to simultaneous parameter variation.~Benchmark 2 is within the variation of Benchmark 1 performed by \cite{Viaux2013} so a data analysis of each model should yield compatible bounds provided that all important uncertainties were accounted for.

\subsection{Grid of Models}

\begin{figure*}[hbp]
    \centering
    \begin{subfigure}
    \centering
        \includegraphics[width=0.8\textwidth]{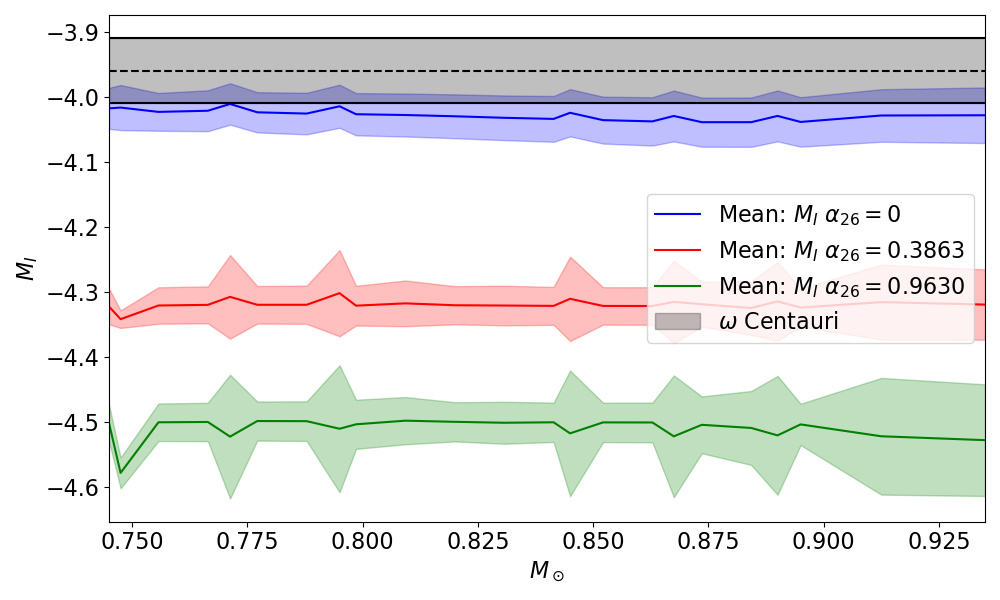}
    \end{subfigure}
    \begin{subfigure}
    \centering
        \includegraphics[width=0.8\textwidth]{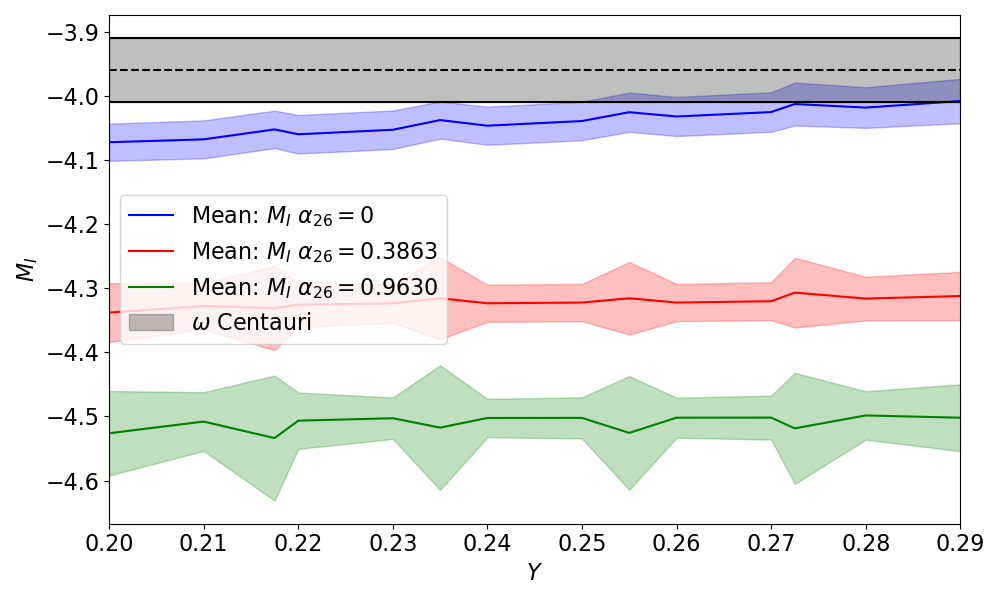}
    \end{subfigure}
\end{figure*}
\clearpage
\begin{figure*}
    \begin{subfigure}
    \centering
        \includegraphics[width=0.8\textwidth]{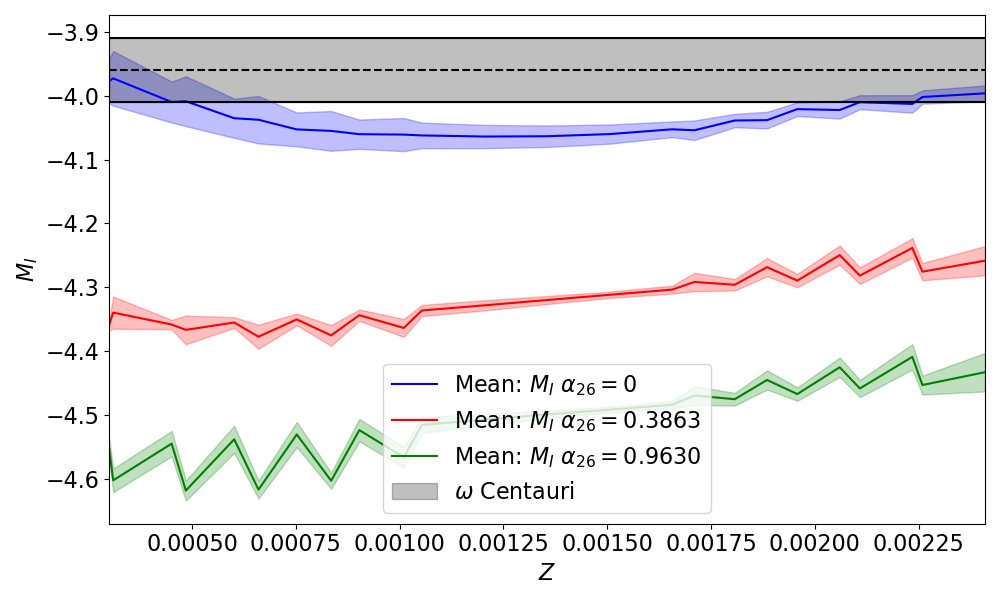}
    \end{subfigure}
    \centering
    \begin{subfigure}
    \centering
        \includegraphics[width=0.8\textwidth]{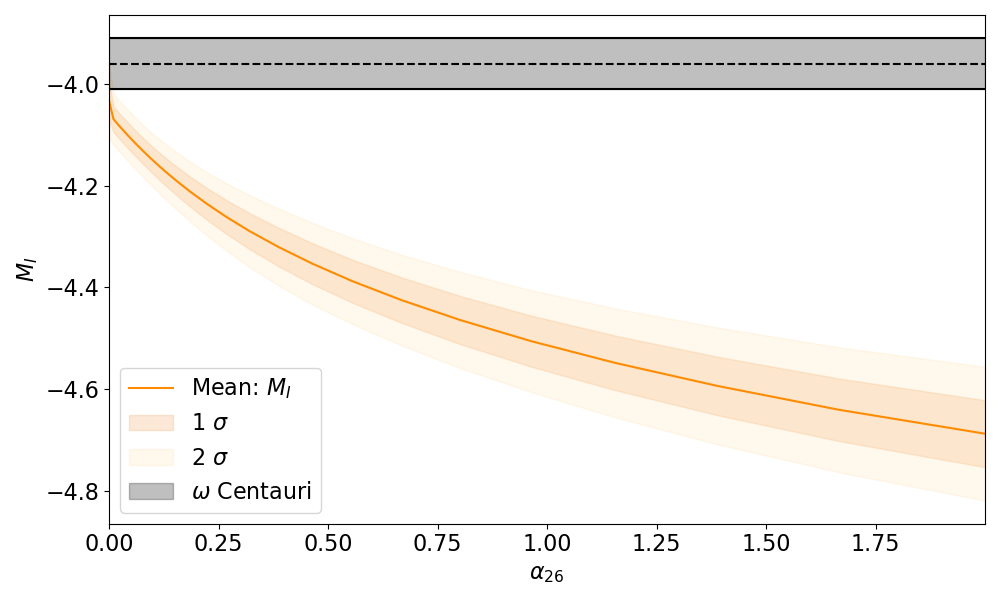}
    \end{subfigure}
    \caption{Variation in the TRGB $M_I$ for Benchmark 1 with varying parameters indicated in the subfigures.~The mean $M_I$ in each bin was found by averaging over the other parameters.~The shaded error bands show the 1$\sigma$ deviation from the mean in each bin.~The dark band represents the calibration of $M_I$ used by \cite{2020ApJ...891...57F}.~The jaggedness is due to the sparser grid spacing outside the nominal prior range (see the main text). We remind the reader that these are the models used to train the ML emulator; physical priors are imposed in the the MCMC analysis that restrict the parameter ranges.~The wider ranges are necessary for efficient training.}
    \label{fig:degeneracy}
\end{figure*}

\begin{figure*}[tbp]
    \centering
    \begin{subfigure}
    \centering
        \includegraphics[width=0.8\textwidth]{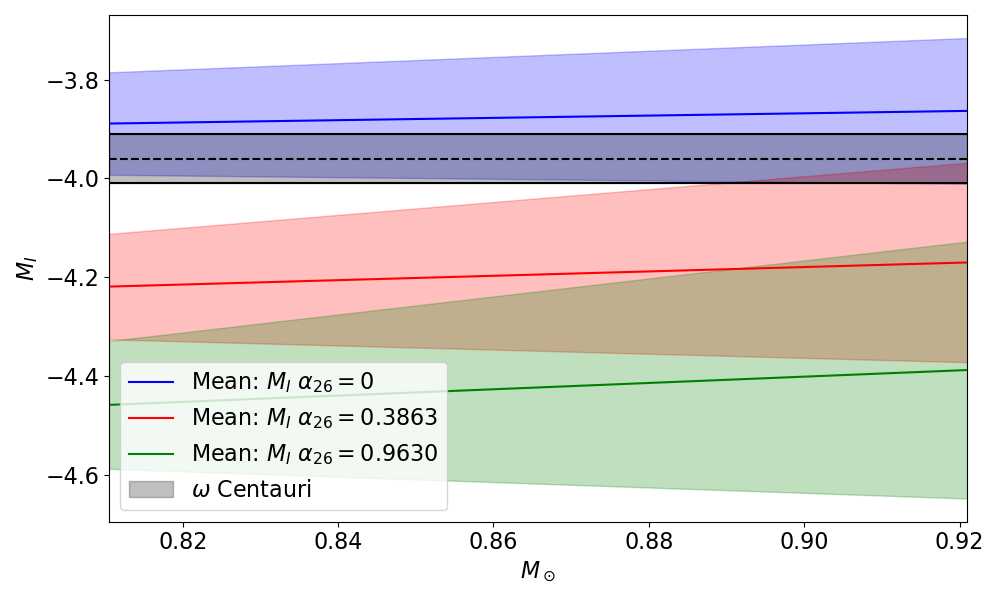}
    \end{subfigure}
    \begin{subfigure}
    \centering
        \includegraphics[width=0.8\textwidth]{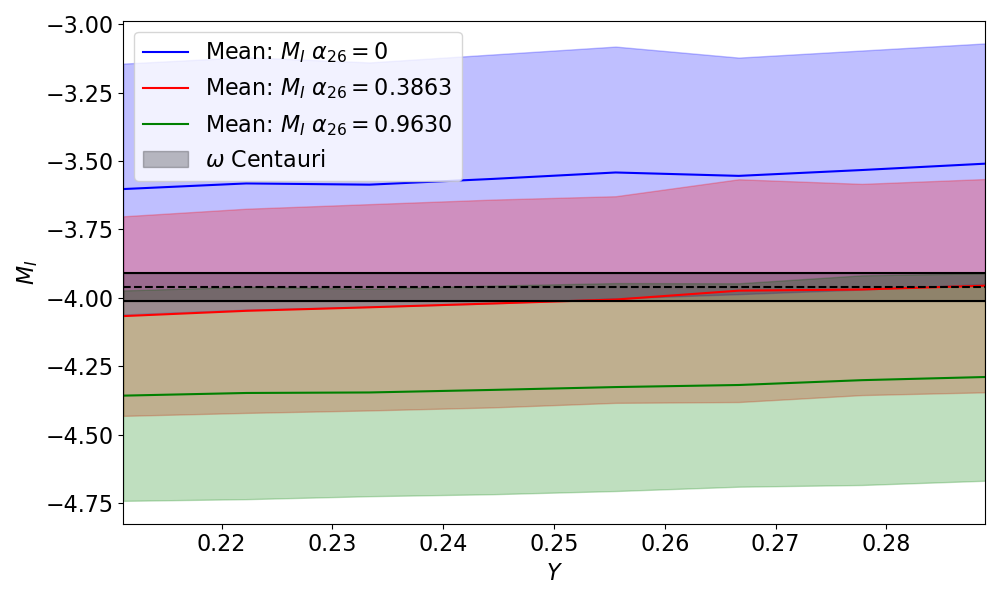}
    \end{subfigure}
\end{figure*}
\clearpage
\begin{figure*}
    \centering
    \begin{subfigure}
    \centering
        \includegraphics[width=0.8\textwidth]{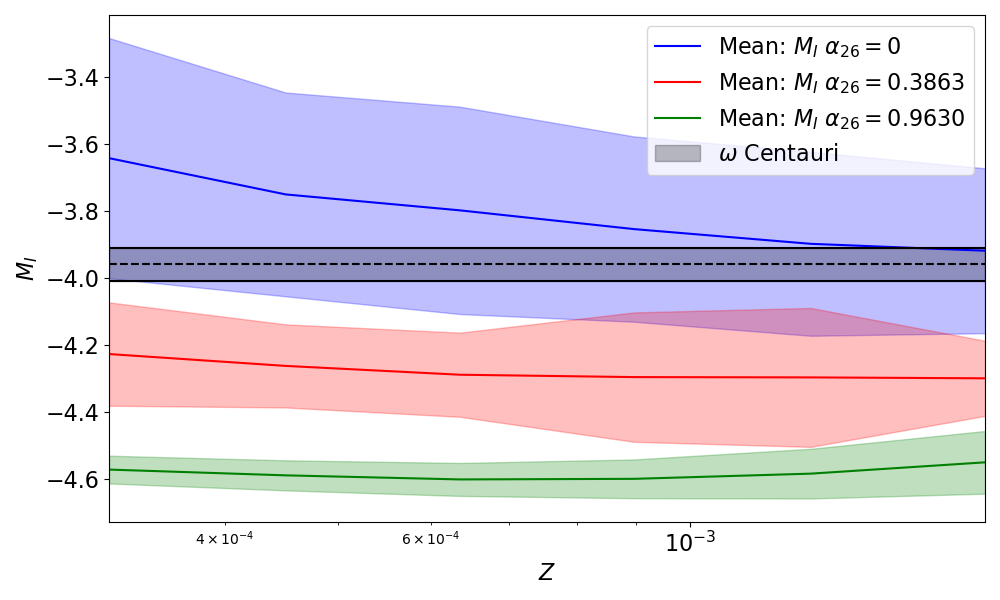}
    \end{subfigure}
    \begin{subfigure}
    \centering
        \includegraphics[width=0.8\textwidth]{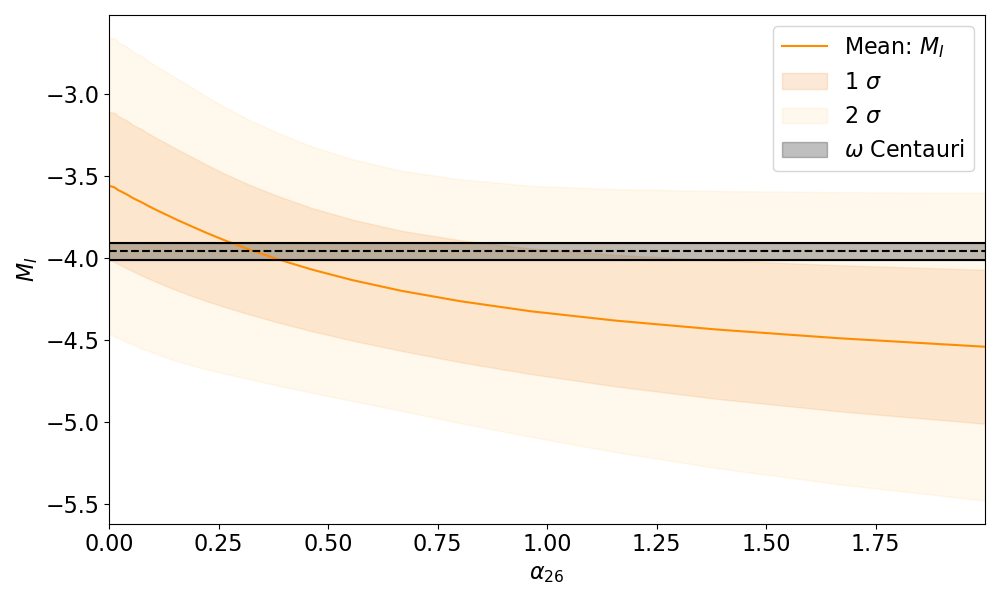}
    \end{subfigure}
    \caption{Same as Figure~\ref{fig:degeneracy} but for Benchmark 2.}
    \label{fig:degeneracy_2}
\end{figure*}

We ran two grids of models for each benchmark given in Table~\ref{tab:benchmarks}.~Each of these is described in turn below. Note that these grids (and their corresponding figures) were generated over a wide range of parameters to enable efficient machine learning. They do not represent the parameters that were ultimately selected by in the MCMC because we imposed physical priors that reduced their ranges.

\textbf{Benchmark 1:}
We evolved models with 15 linearly spaced stellar masses between $0.795$ and $0.845$ $M_{\odot}$, 10 linearly spaced initial helium abundances between 0.235 and $0.260$, 15 linearly initial metallicities between $0.00101$ and $0.00171$, and 30 steps in $\alpha_{26}$ logarithmically spaced between $0.01$ and $2.00$ giving 67,500 models in total.~These regions match the variation given by \cite{Viaux2013}.~We also calculated 18,600 additional models outside these ranges.~We utilized 5 linearly spaced steps in mass from $0.7$ to $0.795$ and an additional 5 linearly spaced steps from $0.845$ to $0.935$, 3 linearly spaced steps in helium from $0.2$ to $0.23$ and an additional 3 linearly spaced steps from $0.26$ to $0.29$, 5 linearly spaced steps in metallicity from $0.00031$ to $0.00101$ and an additional 5 linearly spaced steps from $0.00171$ to $0.00241$, and 30 logarithmically spaced steps from $0.01$ to $2.0$ for $\alpha_{26}$.~These additional models are necessary to capture behavior around the values we use for our priors in our MCMC (the parameter ranges given by \cite{Viaux2013}) in the event the MCMC samples outside those ranges.~However, as this is unlikely, a lower resolution was utilized to preserve computational resources.~For both segments of this grid we added a 31st step, $\alpha_{26} = 0$, to completely represent parameter space.

\textbf{Benchmark 2:}~We ran 116,250 models with varying initial mass $M$, helium abundance $Y$, metallicity $Z$, and $\alpha_{26}$.~The parameters were varied over the ranges $0.7\msun\le M\le2.25\msun$, $0.2\le Y\le  0.3$,  $10^{-5} \le Z\le 0.04$, and $10^{-2} \le \alpha_{26} \le 2$.~These ranges reflect the parameter space that will undergo a core He flash (in $M$, $Y$, and $Z$) \cite{2004sipp.book.....H, 2013sse..book.....K}, and the range where $\alpha_{26}$ has no effect on $M_{I}$ to the edge of where it would begin to affect stars on the main sequence.~We used a linear grid spacing {in mass and helium abundance (15 and 10 steps respectively) and a logarithmic spacing in Metallicity and $\alpha_{26}$ (25 and 30 steps respectively)}.\footnote{This is sufficient to train a ML emulator on our low-dimensional parameter space, but we remark that if one were to vary more parameters then Latin hypercube sampling would be more efficient for such a higher-dimensional space.}~We extended the range of the grid to $\alpha_{26}=0$ by adding 3,750 models from the SM grid simulated by \cite{dennis2025machinelearningtipred}, which was generated in an identical manner to the grid described above i.e, the same version of MESA was used and $M$, $Y$, and $Z$ were varied over the same ranges.~The final grid had 3,750 models per value of $\alpha_{26}$, ensuring that the ML emulator was not biased by disproportional representation of specific values.

Some models in our grids did not reach the TRGB.~These are models that either burn helium stably in the core before executing a helium shell flash (i.e., they do not exhibit a helium core flash and therefore do not contribute to the TRGB), or will not reach the TRGB in the current age of the universe.~A small number of models failed to converge.~These were not numerous enough to affect the ML and were therefore discarded\footnote{The failure to converge is a feature of stellar structure codes and is not due to axions predicting that such stars cannot exist.~One could adjust the numerical solver controls to achieve convergence, but this is inefficient across large grids such as ours and, furthermore, it would be inconsistent to have solver parameters varying across the grid.}.~Discarding these models does not bias the MCMC because the machine learning emulator makes predictions for the corresponding parameters.

The results of our first Benchmark model grid that reach the TRGB in the current age of the universe and execute a core helium flash are compiled together in figure \ref{fig:degeneracy}.~The subfigures show the average value of $M_I$ and its standard deviation for a given bin of varying $M$, $Y$, $Z$, and $\alpha_{26}$.~A band showing the calibration of $M_I$ in $\omega$-Centauri reported by \cite{2001ApJ...556..635B} --- from which \cite{Capozzi2020} derived their strongest bound on $\alpha_{26}$ --- is included for reference.~The figure shows that the variation in $M$, $Y$, and $Z$, is relatively small across this subset of the parameter space, agreeing with previous studies and general expectations.~It also shows that the models do not agree with the empirically-calibrated value of $M_I$ for values of $\alpha_{26} \gtrsim 0.6$  within $2\sigma$ when degeneracies across other parameters are not accounted for. 

Figure~\ref{fig:degeneracy_2} shows the results for our second Benchmark grid.~Important differences between figures \ref{fig:degeneracy} and \ref{fig:degeneracy_2} are evident.~Most importantly, a broader range of $\alpha_{26}$ (now the entire parameter range) agrees with the value of $M_I$ obtained by \cite{Capozzi2020}. We also see that unlike for figure \ref{fig:degeneracy}, where almost exclusively $\alpha_{26} = 0$ is the only value of agreement for the $M$, $Y$, and $Z$ parameters, figure \ref{fig:degeneracy_2} has a larger range of $\alpha_{26}$ that are compatible with a range of values for the $M$, $Y$, and $Z$ parameters. 

\begin{figure*}[ht]
    \centering
    \begin{subfigure}
        \centering
        \includegraphics[width=0.48\textwidth]{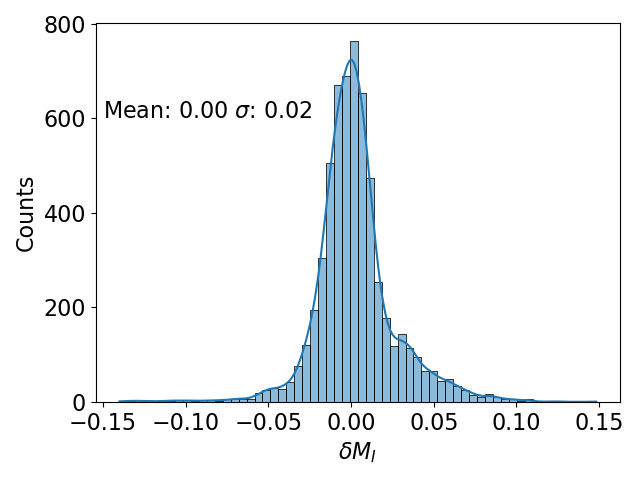}
    \end{subfigure}
    \begin{subfigure}
        \centering
        \includegraphics[width=0.48\textwidth]{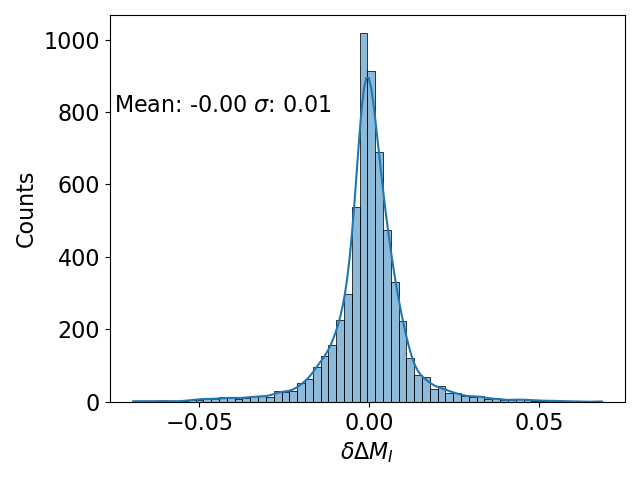}
    \end{subfigure}
    \begin{subfigure}
        \centering
        \includegraphics[width=0.48\textwidth]{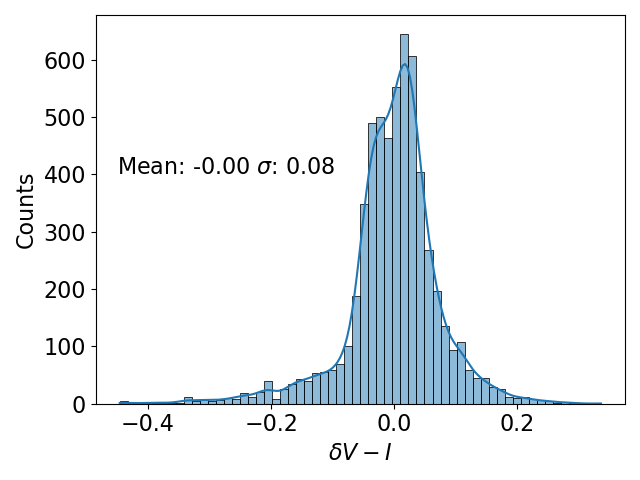}    
    \end{subfigure}
    \begin{subfigure}
        \centering
        \includegraphics[width=0.48\textwidth]{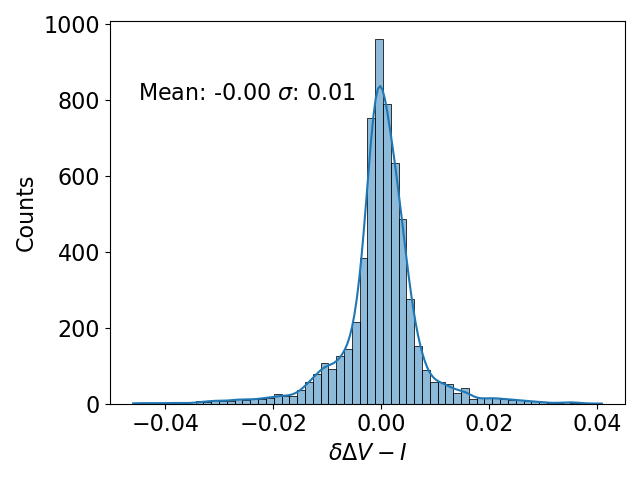}
    \end{subfigure}
    \caption{Error distributions for $M_I$, $\Delta M_I$, $(V-I)$, and $\Delta (V-I)$ from the ML regression for  Benchmark 1.~These distributions were calculated by subtracting the value of the ML prediction for each quantity  from the value found by applying the WL code directly to the MESA outputs for each point in our grid.}
\label{fig:mError_Standard}
\end{figure*}
 
\begin{figure*}[ht]
    \centering
    \begin{subfigure}
        \centering
        \includegraphics[width=0.48\textwidth]{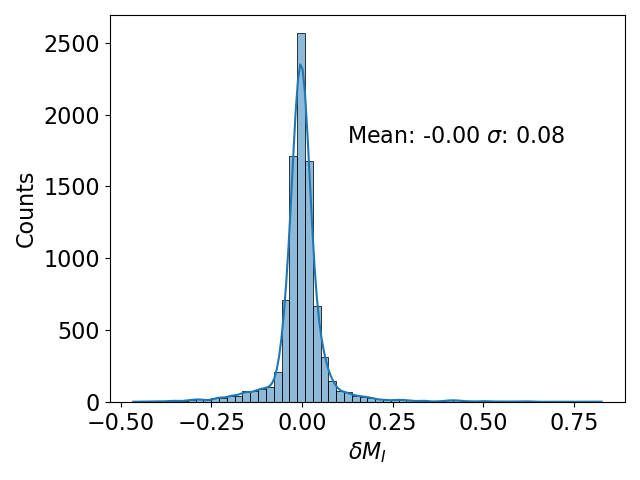}
    \end{subfigure}
    \begin{subfigure}
        \centering
        \includegraphics[width=0.48\textwidth]{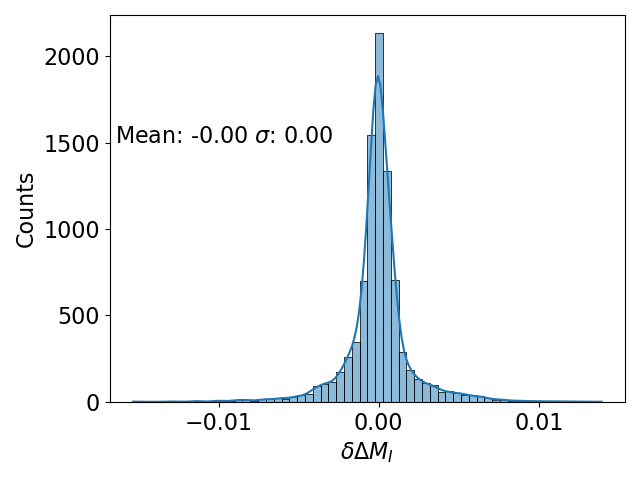}
    \end{subfigure}
    \begin{subfigure}
        \centering
        \includegraphics[width=0.48\textwidth]{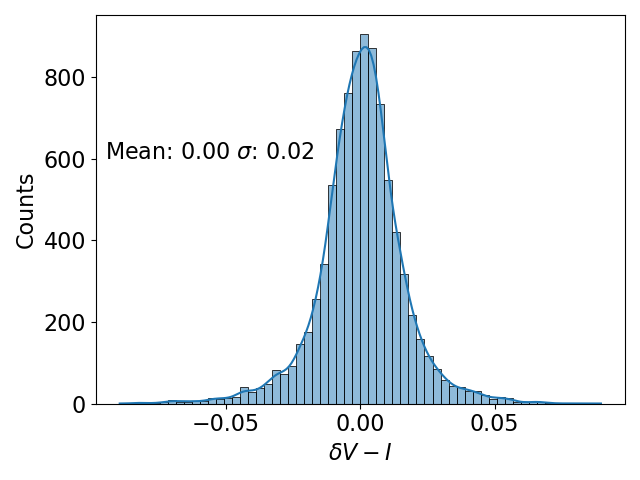}
    \end{subfigure}
    \begin{subfigure}
        \centering
        \includegraphics[width=0.48\textwidth]{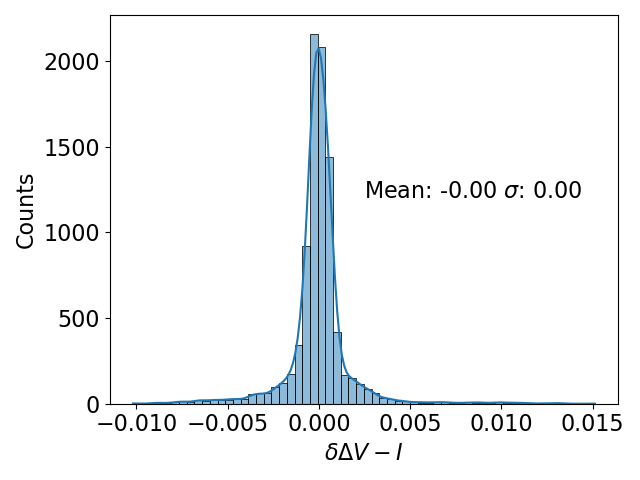}
    \end{subfigure}
    \caption{The same as \ref{fig:mError_Standard} but for Benchmark 2.}
\label{fig:mError}
\end{figure*}

\section{Machine Learning Emulator}
\label{sec:ML}
For both Benchmark models in Table~\ref{tab:benchmarks}, we trained separate two component deep neural network (DNN) emulators for the individual model grids described above using tensorflow \cite{tensorflow2015-whitepaper} and keras \cite{chollet2015keras}.~All emulator algorithm components require the grid parameters $\{M,Y,Z,\alpha_{26}\}$ as their inputs.~The first component of both emulators was a classifier which accepted models which successfully reached the TRGB, and flagged those that either exhibited a helium shell flash or did not reach the TRGB in the current age of the universe.~The second component was a regression algorithm that predicts $M_I$, $\Delta M_I$, $V-I$, and $\Delta V-I$.~Additional regressors for each Benchmark are included in our reproduction package.~These were trained to predict luminosity $L$, effective temperature $T_{\rm eff}$, surface gravity $\log_{10}(g)$, and iron abundance $[{\rm Fe}/{\rm H}]$.~These algorithms were less accurate than emulating the magnitude directly.~We include them for those interested in using different bolometric corrections.

We built our DNNs with an ADAM \cite{2014arXiv1412.6980K} optimizer and hand tuned our hyperparameters to optimize the network training.\footnote{We remark that our success with hand tuning may not be replicated if the number of model parameters is increased, and that it may be necessary to use methods such as grid search, random search, or the genetic algorithm to tune hyperparameters as explored in \cite{2019arXiv191206059L}.}~All algorithms were trained using an 80\%/10\%/10\% split between training, validation, and testing data.~However, for both Benchmarks the classifier training set was resampled using the Synthetic Minority Oversampling Technique (SMOTE) \cite{2011arXiv1106.1813C} to rebalance the training data which results in a more accurate emulator \cite{2019SPIE11198E..13S}.~An unbalanced dataset can bias the network to labelling more objects as the more populous class while achieving similar levels of accuracy.

For Benchmark 1, the classification algorithm has an accuracy of 98.0\% and a cross-entropy loss of 0.055.~The regression algorithm predicts $M_I$, $\Delta M_I$, $(V-I)$, and $\Delta (V-I)$ with a mean squared error loss of $0.001596$ for input and output data that has been normalized between 0 and 1.~Even though this value is larger than for our second Benchmark, this model is more accurate when the root mean squared errors of the denormalized data are compared.~The normalizing process can influence the value of the loss if there are outliers at low or high values that artificially compress the majority of the data into a narrow range near 1 or 0 respectively.~As in figure \ref{fig:mError}, the histograms for the residuals of this emulator are presented in figure \ref{fig:mError_Standard}.~These errors are subdominant to the errors from the bolometric corrections (the center of the distribution is essentially zero).

For Benchmark 2, the classification algorithm has an accuracy of 99.2\% and a cross-entropy loss of 0.023.~The regression algorithm predicts $M_I$, $\Delta M_I$, $(V-I)$, and $\Delta (V-I)$ with a mean squared error loss of $2.115 \times 10^{-5}$ for input and output data that has been normalized between 0 and 1.~Histograms of the residuals from the testing data for $M_I$, $\Delta M_I$, $V-I$, and $\Delta(V-I)$ are given in figure \ref{fig:mError}.~The residual plots show that our errors are similarly subdominant to the errors from the bolometric corrections, which have an average error of $\sim0.1$ mag.

\section{MCMC}
\label{sec:MCMC}

In this section, we  attempt to constrain $\alpha_{26}$ for each of our benchmark models by using an MCMC sampler to compare the theoretical value of $M_I$ predicted by our ML emulator trained in section \ref{sec:ML} with the calibrated values in the LMC, NGC 4258, and $\omega$-Centauri (both values) listed in table \ref{tbl:calibrations}.~Our analyses employed the \textit{emcee} package \cite{2013PASP..125..306F}.~

\subsection{Priors}

For both our benchmarks, physical priors on $M$, $Y$, and $Z$ were needed to place bounds on $\alpha_{26}$, as it is clear from Figures \ref{fig:degeneracy} and \ref{fig:degeneracy_2} that a broad range of $\alpha_{26}$ is possible from unconstrained $M$, $Y$, and $Z$.~We used the same priors as \cite{Viaux2013, 2013PhRvL.111w1301V,Capozzi2020} in order to mirror their analyses as closely as possible given the different methods of uncertainty accounting --- single parameter variation vs.~MCMC.~Specifically, we used Gaussian priors $M=0.820\pm0.025{\rm M}_\odot$, $Y=0.245\pm0.015$, and $Z=0.00136\pm0.00035$ and a uniform prior on $\alpha_{26}$ over the range $0 < \alpha_{26} \le 2$.~The stellar priors correspond to the model for the globular cluster M5 introduced by \cite{Viaux2013}.~The metallicity was derived from observations of the iron abundance and the helium mass fraction was derived from observations of extragalactic HII regions.~We have updated this to impose the lower bound $Y\ge 0.245$ reported by the more recent Planck observations \cite{Planck:2018vyg}.~The mass prior was chosen so that the age of M5 is $13.8$ Gyr given their adopted chemical composition and its uncertainties.~We note that with MCMC it is possible to vary age directly instead of using the mass as a proxy, but we have not done this in order to perform a fair comparison with previous works.

\subsection{Likelihood}
In both cases, the MCMC used the ML classification algorithm to assign a zero likelihood to models which do not successfully core helium flash within the age of the universe.~This procedure ensures that regions of parameter space corresponding to stars that don't contribute to the TRGB are not explored, and can be thought of as imposing a prior that enforces our knowledge of the regions of parameter space already excluded on theoretical grounds.~The log-likelihood function was taken to be Gaussian i.e., 

\begin{equation}
    \ln{\mathcal{L}} = -\frac{1}{2}\left[\left(\frac{M_{I, \textrm{obs}} - \textrm{ML}(\theta)_{I})^2}{\sigma_{\textrm{tot}}}\right)^2 \right.~+ \left.~\ln(2\pi \sigma_{\textrm{tot}}^2)\vphantom{\sum^N_{i=1}}\right]
    \label{eq:MLE},
\end{equation}

where $M_{I, \textrm{obs}}$ is one of the observed I Band values from table \ref{tbl:calibrations} and ML$(\theta)_{I}$ is the ML prediction for $M_I$ for a given set of parameters $\theta$, and ~$\sigma_{\textrm{tot}}$ is given by
\begin{equation}
\sigma_{\textrm{tot}}^2 = \sqrt{\sigma_{\textrm{obs}}^2 + ML(\theta)_{\Delta M_I}^2},
\end{equation}

where $\sigma_{\textrm{obs}}$ is the error on the empirical I band calibration and $ML(\theta)_{\Delta M_I}$ is the Worthey \& Lee bolometric error predicted by the ML.~We accounted for the error in the ML by randomly drawing from the distributions in Figure \ref{fig:mError} and adding them to the predictions from the ML emulator for each $\theta$, similar to the procedure in \cite{2019MNRAS.482.1352M}.~These errors are subdominant to the errors on the calibration and the bolometric correction, but were included for completeness.

The error accounting above is similar to that of \cite{Capozzi2020}. However, where they add in quadrature the adopted uncertainties from \cite{Serenelli2017} and \cite{Viaux2013}, we add our uncertainties at each step of the MCMC. While we use the same observational uncertainties as \cite{Capozzi2020} given in their Table 1, our bolometric correction uncertainty is predicted for a given set of parameters by our machine learning as described in our ML section and added in quadrature to the observational uncertainty. This is in contrast to simply adopting a single error for the bolometric correction which \cite{Viaux2013} state as a maximum range of uncertainty. Our method accounts for the fact that the bolometric correction is parameter dependent, and therefore our bolometric error will be more accurate and likely smaller on an individual model basis than the works that assume a single maximum error. Further, \cite{Viaux2013} and others add the errors for M, Y, and Z in quadrature while we account for them using the simultaneous variation of the MCMC.  Lastly, the integral performed by \cite{Capozzi2020} to calculate the bound on $\alpha_{26}$ is equivalent to the resulting posterior from our MCMC analysis. We do introduce one additional source of error not present in previous works: the systematic error introduced by the machine learning. This is accounted for in our procedure described above, and is subdominant to the other errors in the analysis. Therefore it does not affect the acquired bound.

For each MCMC run, we determined that the chains had converged when the integrated autocorrelation time $\tau$ was less than 0.1\% the length of the chain and that it had changed by less than 1\% over the previous 10,000 points \cite{2010CAMCS...5...65G,Sokal}.~We found that both MCMCs converged within 120,000 steps but we allowed them to continue to 500,000 to ensure that the walkers were no longer influenced by their starting location.~We discarded half of the samples as burn-in.

\section{Results}

We performed MCMC analyses using empirical TRGB $M_I$ calibrations in the LMC \cite{2019ApJ...886...61Y} for Y19 and \cite{2020ApJ...891...57F} for F20;~NGC 4258 \cite{2017ApJ...835...28J};~and $\omega$-Centauri \cite{2001ApJ...556..635B}.~We use the updated values for NGC 4258 and $\omega$-Centauri from \cite{Capozzi2020} to directly compare with their results.~We also present results for an updated value $M_I$ from $\omega$-Centauri calculated using the procedure of \cite{Capozzi2020} applied to the recently updated distance from \cite{Baumgardt2021}.~The corner plots for the other systems (and the updated value for $\omega$-Centauri) we studied for both benchmarks with Gaussian priors are visually similar to those shown for \cite{Capozzi2020}'s $\omega$-Centauri and are included in in Appendix \ref{sec:other_MCMC}.

\begin{figure}
    \centering
    \includegraphics[width=\textwidth]{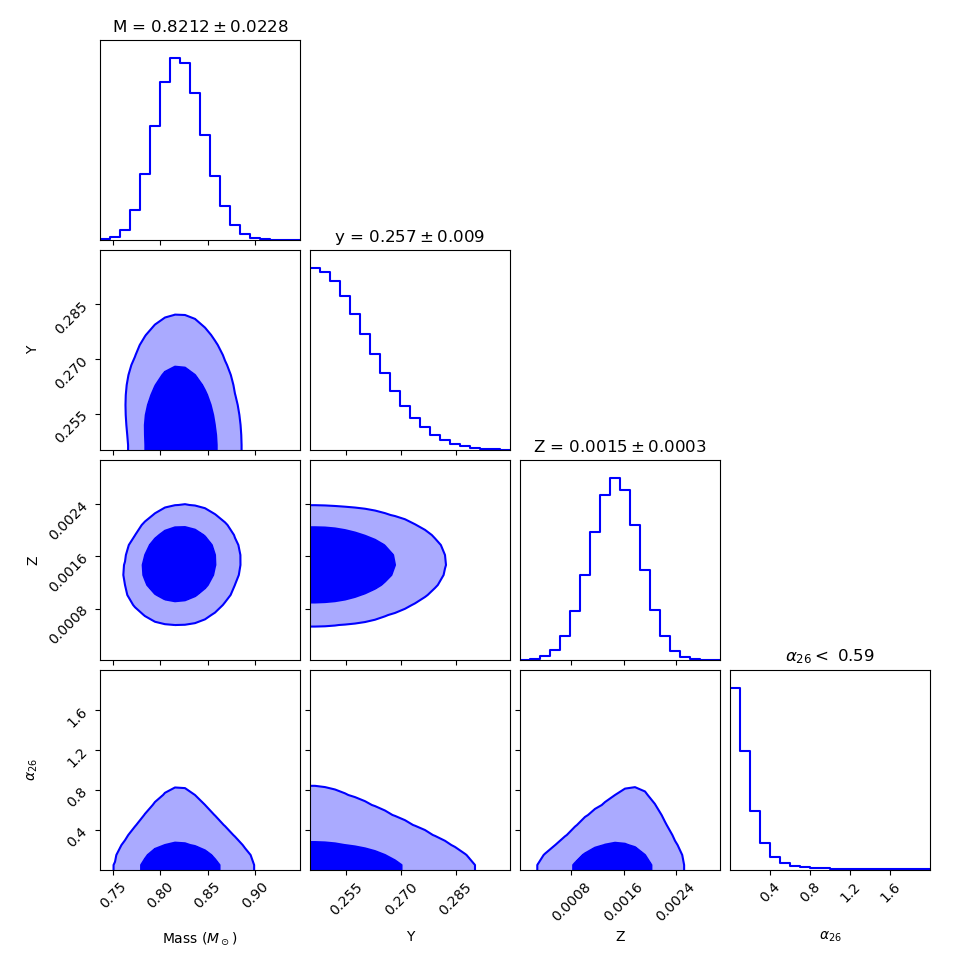}
    \caption{Posterior distributions for each parameter for the $\omega$-Centauri calibration of $M_I$ assuming Benchmark 1.~Both the 2D and marginalized posteriors are shown.~The titles on the marginalized posteriors represent the means and standard deviations for $M$, $Y$, and $Z$.~For $\alpha_{26}$, the title represents the 95\% confidence limit.~The contours represent the 68\% and 95\% confidence intervals.}
\label{fig:cornerTightPriorsStandard1}
\end{figure}

\begin{figure*}
    \centering
    \includegraphics[width=\textwidth]{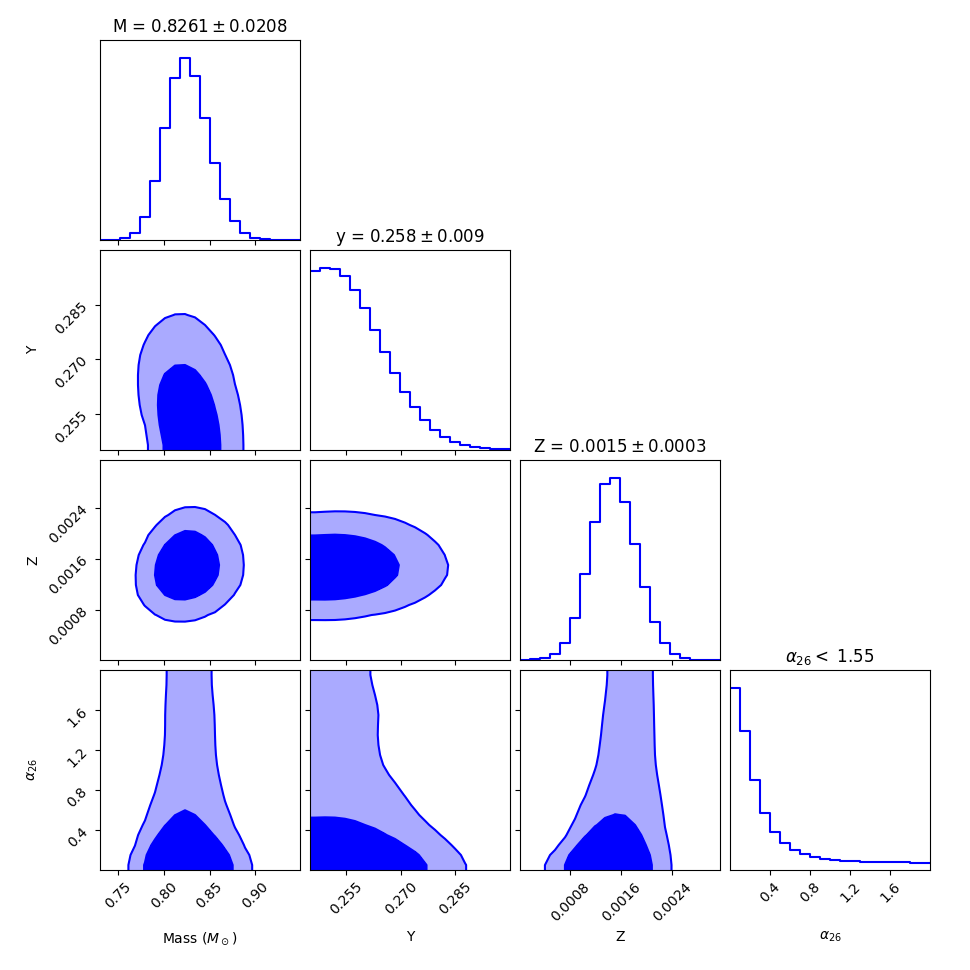}
    \caption{Same as Fig.~\ref{fig:cornerTightPriorsStandard1}, but for Benchmark 2.}
    \label{fig:cornerTightPriors1}
\end{figure*}

The results for $\omega$-Centauri for Benchmark 1, which we remind the reader is similar to the benchmark model of reference \cite{Viaux2013}, are shown in figure \ref{fig:cornerTightPriorsStandard1}.~The priors for $M$, $Y$, and $Z$ are completely recovered by the MCMC, and we find a constraint on the axion-electron coupling $\alpha_{26} < 0.59$ at the 95\% confidence level, a factor of four larger than the bound $\alpha_{26}<0.2$ found by \cite{Capozzi2020} for the same system.~The results for Benchmark 2, which we remind the reader uses different values of the mass loss and mixing length within the variation performed by \cite{Viaux2013}, are similarly shown in figure~\ref{fig:cornerTightPriors1}.~The 95\% limit falls at $\alpha_{26}<1.55$, greater than an order-of-magnitude weaker than reported by \cite{Capozzi2020}.~The posterior for $\alpha_{26}$ does not decay to zero at $2.0$, the edge of our prior range, suggesting that the bound may be weaker than we report.~We did not have the computing power to extend the range of $\alpha_{26}$, so were unable to explore this further in this work.~We discuss this in our conclusions below (Sec.~\ref{sec:conc}).

\begin{figure*}
    \centering
    \includegraphics[width=0.85\textwidth]{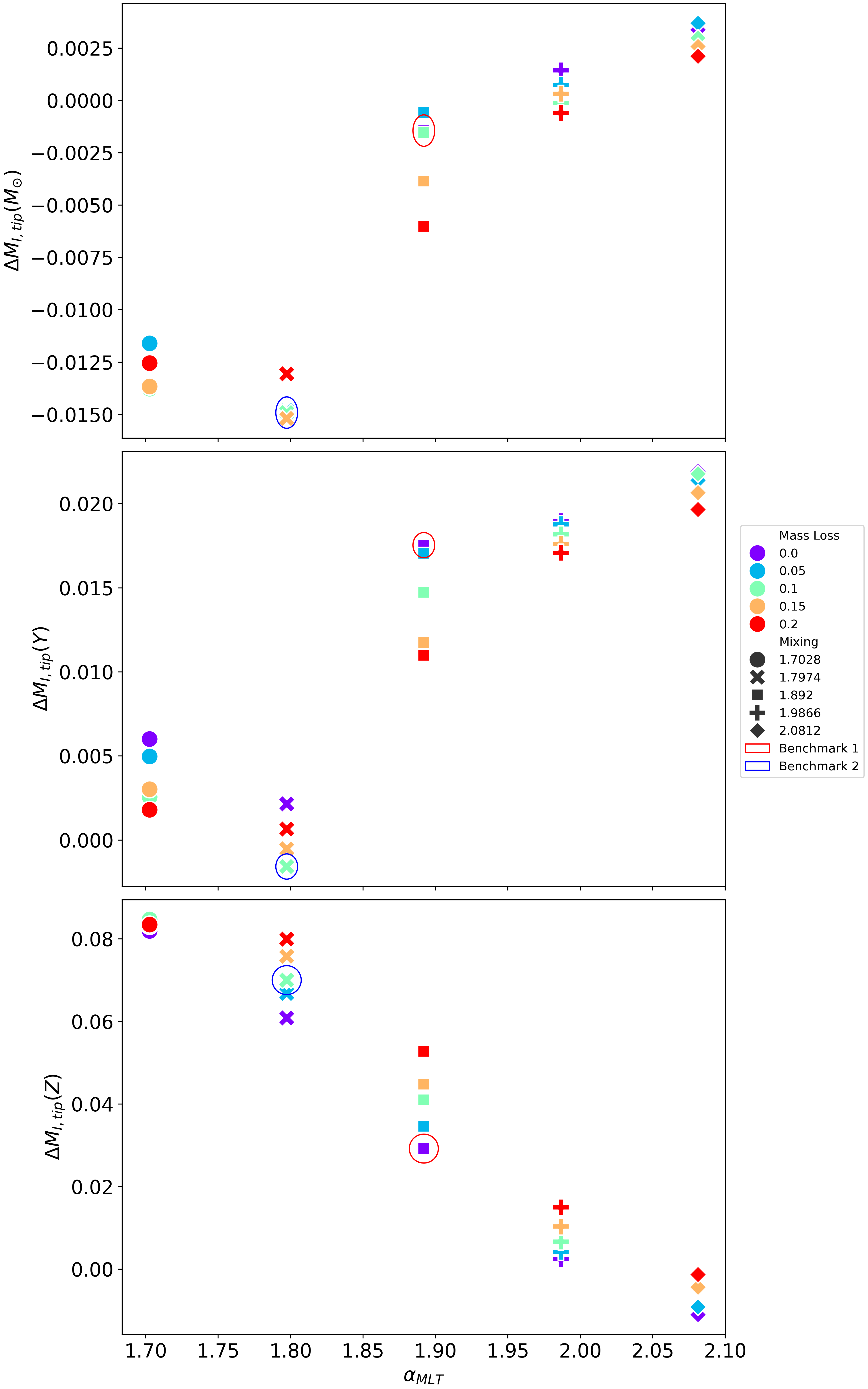}
    \caption{Variation in $M_I$ for the sparse grids that varied initial mass, helium, metallicity, mixing length, and wind loss for the SM.~We plotted the $\Delta M_{I, tip}$ that resulted from varying $M$ (top plot), $Y$ (middle plot), and $Z$ (bottom plot) across their prior ranges.~The marker shapes indicate changes in mixing length, and the marker color indicate changes in mass loss efficiency.~Note that Benchmark 1 corresponds to the purple square and Benchmark 2 to the green X.~The figure shows that the spread in $\Delta M_{I, tip}$ is strongly dependent on mixing length and $\alpha_{26}$ parameter, and weakly dependent on mass loss.}
    \label{fig:derivative_plot_SM}
\end{figure*}

\begin{figure*}
    \centering
    \includegraphics[width=0.85\textwidth]{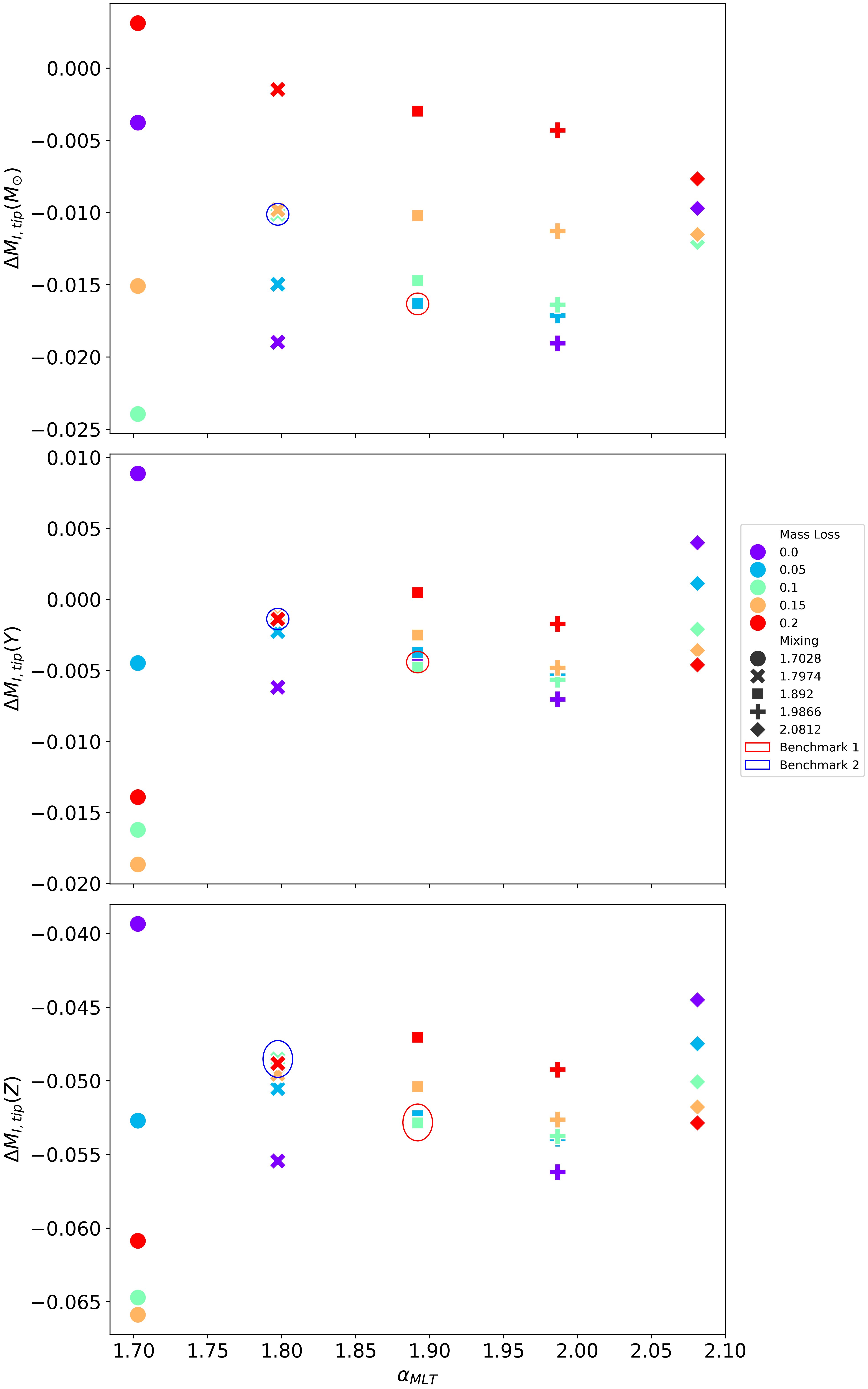}
    \caption{The same as \ref{fig:derivative_plot_SM}, but for $\alpha_{26} = 0.7$}
    \label{fig:derivative_plot_DM}
\end{figure*}

The two benchmarks differ in their adopted mixing length and mass loss efficiency parameters, and  variances in $\Delta M_{I, tip}$ (the difference in the $M_I$ at the TRGB for similar input parameters) induced by varying these two parameters were accounted for by \cite{Viaux2013,2020PhRvD.102b3007D}.~Covariances were not accounted for so to investigate them as a potential origin for the discrepancy we ran a grid of models with varying stellar mass, helium mass fraction, metallicity, mixing length parameter, and wind loss efficiency for both the SM and $\alpha_{26}=0.7$;~and calculated $\Delta M_{I, tip}$ by varying $M$, $Y$, and $Z$ across their prior ranges.~The results are shown in Figures \ref{fig:derivative_plot_SM} and \ref{fig:derivative_plot_DM} for the SM ($\alpha_{26} = 0$) and $\alpha_{26} = 0.7$, respectively.~Evidently, $\Delta M_{I, tip}$ is a strongly varying function of mass loss, mixing length, and $\alpha_{26}$.~The values of $\Delta M_{I, tip}$ for Benchmark 1 with $\alpha_{26}=0$ are commensurate with those reported by \cite{Viaux2013} while those for Benchmark 2 are larger by a factor of two or more.~We therefore conclude that the different bounds obtained are the result of covariances between both stellar input physics and $\alpha_{26}$ not accounted for by single-parameter variation.

\section{Conclusions}
\label{sec:conc}
In this work we have presented a novel method for incorporating uncertainties and degeneracies into stellar tests of physics beyond the standard model.~Focusing on tip of the red giant branch bounds on the axion-electron coupling found using the I band magnitude $M_I$, we simulated grids of stellar models with varying initial mass $M$, helium abundance $Y$, metallicity $Z$, and axion-electron coupling $\alpha_{26}$.~Due to limited computing power, we fixed other input physics to fiducial values.~We simulated two grids with different values of mass loss efficiency and mixing length with all other input physics fixed.~One of these benchmarks was similar to the fiducial model used for previous analyses by \cite{Viaux2013,Capozzi2020} that employed single parameter variation to estimate the uncertainties, and the other lies within the variation performed as part of that study.~We trained  machine learning emulators on each grid to predict the I band magnitude (and errors due to bolometric corrections) as a function of $\{M,Y,Z,\alpha_{26}\}$.~This was then used in Markov Chain Monte Carlo analyses to compare with empirical calibrations and derive new bounds that incorporate the effects of covariances between parameters.~The novelty of our method lies in substituting the stellar structure code with our emulator in the MCMC.~The long run-times of stellar structure codes (hours) is a major barrier to using MCMC to compare their predictions with data.~Our emulator evaluates in milliseconds.~Using MCMC enabled us to vary uncertain stellar modeling parameters simultaneously to assess the importance of degeneracies for the first time.~

Application of our pipeline to our two Benchmarks yielded bounds of $\alpha_{26}<0.75$ for Benchmark 1 and $\alpha_{26}<1.58$ for Benchmark 2.~These two values differ by a factor of  $\sim$2~The first Benchmark yielded a bound weaker than previous results by nearly a factor of four while the value resulting from the second Benchmark was weaker by a factor of almost 8.~We investigated the origin of the discrepancy, and traced it to covariances between the stellar parameters, finding that the variation in the TRGB I band magnitude with $\{M,Y,Z\}$ is a strongly varying function of mass loss and mixing length.

Our results suggest that bounds on the axion-electron coupling obtained by comparing theoretical and calibrated TRGB I band magnitudes (e.g. \cite{Viaux2013, Capozzi2020}) may be overestimated and that the bounds obtained by these methods are sensitive to the choice of input physics due to covariances between parameters.~In addition, the posterior for our second benchmark did not fully decay to zero before reaching the upper limit on $\alpha_{26}$, indicating that our bound may itself be over-estimated.~We did not have the computing power to extend our training grid beyond $\alpha_{26}=2.0$ but our results suggest that future work doing so would be valuable.

We note that our conclusions do not fully extend to the alternate method where observations of the bolometric magnitude $M_{\rm bol}$ are compared with theoretical predictions \cite{Straniero2020,Troitsky:2024keu} because this method does not require the effective temperature to compute the bolometric corrections, so is not subject to all of the uncertainties we have studied here.~

In the remainder of this section, we discuss limitations of our present study and how they could be overcome to achieve this goal, and potential applications of our method to other astrophysical probes of physics beyond the Standard Model.

\subsection{Limitations of Our Study}
Our work is a preliminary study and, as such, is subject to some caveats and limitations.~First, we only varied the mass, initial helium abundance, and initial metallicity of the stars we simulated;~we held the stellar input physics such as nuclear reaction rates and neutrino energy loss rate fixed to fiducial values.~These are known to be equally large sources of uncertainty \cite{Serenelli2017,2022MNRAS.514.3058S}.~We made this choice for two reasons.~The first is computational resources.~Our grid for Benchmark 2 (our larger grid) took 1.23 million CPU hours to complete.~Adding variation in the mass loss, mixing length, nuclear reaction rates (there are two important rates), neutrino loss rates, in addition to the four parameters we already varied ($M$, $Y$, $Z$, and $\alpha_{26})$ (nine parameters in total) would take approximately 2.6 billion CPU hours.~More efficient algorithms for sampling parameter space e.g., Latin hypercube sampling or active learning \cite{2022arXiv220316683A} would aid in decreasing the number of models needed per parameter.~Second, it is possible that the accuracy of the ML emulator will be degraded by adding additional parameters.~It is important that the errors in our constraint (or lack thereof) is due to experimental errors and not the ML error.~For this reason, we chose to focus on the minimal number of parameters needed to produce physically reasonable variations\footnote{Note that $M$, $Y$, and $Z$ vary between objects and so, even if the stellar input physics parameters were known to infinite precision, there would still be variation in the TRGB $M_I$ due to these parameters.} to ensure an accurate emulator.

Another caveat is that we have made the assumption that certain model parameters, (e.g. $\alpha_{\textrm{MLT}}$) remain constant throughout the evolution of a star. This is not guaranteed to be the case as shown by \cite{Ferraro2006, Salaris1996}. Changes in parameters along the stellar evolutionary track that affect the input physics can have significant effects on the following stages of evolution of a star and further study is required to understand the potential impact of these parameter fluctuations on measurements of the TRGB. This assumption was also made by \cite{Viaux2013, Serenelli2017, Capozzi2020} and others, and therefore we adopted it as well.

The third caveat is that we have assumed that the TRGB I band magnitude is due solely to the brightest star.~In practice, $M_I$ is calibrated using edge detection techniques (e.g., \cite{2018A&A...615A..96M}) applied to the color-magnitude diagram.~We made this choice to enable a comparison with previous works, which make the same assumption \cite{1995PhRvD..51.1495R,Capozzi2020}.~We note that our emulator could be used to make theoretical predictions found using the same method as the empirical calibration.~In particular, one could use our emulator to simulate a mock color-magnitude diagram by drawing $M$, $Y$, and $Z$ from some reasonable distribution for the specific system under study then apply the same edge-detection techniques to extract $M_I$ as a function of $(V-I)$ i.e., the zero-point and color-correction can be theoretically-predicted.~One could even MCMC over these mock diagrams.

The final caveat is that our analysis includes statistical errors but not systematic errors.~We have made specific choices for the input physics, including discrete choices such as initial elemental abundances which affect the theoretical prediction for $M_I$ via its dependence on [Fe/H]), choice of bolometric corrections, and choice of stellar structure code.~Furthermore,  MESA may be missing physics such as three-dimensional processes.~All of these act as a source of systematic uncertainty.~Different choices of code, input physics, or bolometric corrections will not alter our conclusions.~Changing code and/or input physics will simply result in a systematic offset of the best-fitting parameters, and other bolometric corrections have errors comparable to those we adopted in this work so will not alter the uncertainties we derived.~It would be interesting to investigate the effects of varying these choices.

\subsection{Application to Other Stellar Tests of New Physics}
All astrophysical systems are subject to degeneracies and uncertainties.~Our work has highlighted the paramount importance of fully accounting for these covariances when using astrophysical systems as probes of physics beyond the Standard Model.~The methods we have developed here could be applied to reevaluate the bounds obtained using other stellar tests of axions e.g., horizontal branch stars \cite{Ayala:2014pea,Carenza:2020zil} the white dwarf luminosity function \cite{MillerBertolami:2014rka,Dolan:2021rya}, pulsating white dwarfs \cite{Corsico:2019nmr}, black hole population statistics \cite{2021PDU....3200801C,2020arXiv200707889C,Sakstein:2020axg,Straight:2020zke,Baxter:2021swn,Sakstein:2022tby,2022arXiv220801110C,Croon:2023trk}, and Cepheid stars \cite{Friedland:2012hj}.~Another  application would be to reevaluate the stellar bounds on other new physics models such as hidden photons, and theories where the neutrino has a large magnetic dipole moment \cite{2023arXiv230713050F}.

\section*{Acknowledgments}
We thank Adrian Ayala, Aaron Dotter, Robert Farmer, Frank Timmes, and the wider MESA community for answering our MESA-related questions.~We are grateful for conversations with Eric J.~Baxter, Djuna Croon, Samuel D.~McDermott, Harry Desmond, Noah Franz, Marco Gatti, Dan Hey, Esther Hu, Jason Kumar, Danny Marfatia,  Marco Raveri, David Rubin, Xerxes Tata, Brent Tully, and Guy Worthey.~We are especially thankful to David Schanzenbach for his assistance with using the University of Hawai\okina i MANA and KOA supercomputers.

Our simulations were run on the University of Hawai‘i’s high-performance supercomputers MANA and KOA.~The technical support and advanced computing resources from University of Hawai‘i Information Technology Services – Cyberinfrastructure, funded in part by the National Science Foundation MRI award \#1920304, are gratefully acknowledged.

\section*{Software}
MESA version 12778, MESASDK version 20200325, Worthey and Lee Bolometric Correction Code \cite{WortheyLee2011}, NumPy version 1.22.3 \cite{2020Natur.585..357H} \texttt{Pandas} version 1.4.3 \cite{pandasDataStructure, pandasSoftware}, \texttt{Matplotlib} version 3.5.1 \cite{2007CSE.....9...90H}, \texttt{TensorFlow} version 2.4.1 \cite{tensorflow2015-whitepaper}, \texttt{Keras} version 2.8.0 \cite{chollet2015keras}, \texttt{corner} version 2.2.1 \cite[][]{corner}, \texttt{emcee} version 3.1.2 \cite{2013PASP..125..306F}.

\appendix
\section{Axions and Implementation into MESA}
\label{sec:axions}

The Lagrangian for an axion-like particle $a$  is
\begin{align} \label{eq:axion_lag}
    \mathcal{L} = - i g_{ae} a \bar\psi_e \gamma_5 \psi_e  - \frac12 m_a^2 a^2,
\end{align}
where $g_{ae}$ is the (dimensionless) axion-electron coupling, $\psi_e$ is the electron's Dirac spinor field, and we have neglected the axion-photon coupling since its effects are negligible inside TRGB stars \cite{Altherr:1993zd}.~It is common to work with the $\mathcal{O}(1)$ quantity $\alpha_{26} = 10^{26} g^2_{ae}/4\pi$, which we adopt in this work.~Other axion couplings to standard model fields are allowed but are not relevant for this study.

{The} rate of energy loss per unit time per unit mass due to the interaction in equation \eqref{eq:axion_lag} is \cite{1995PhRvD..51.1495R}
\begin{equation}
    Q_{ae} = Q_{sC} + (Q_{b, ND}^{-1} + Q_{b, D}^{-1})^{-1},
    \label{eq:loss}
\end{equation}
where $Q_{sC}$ is the loss from semi-Compton scattering, $Q_{b, ND}$ is the non-degenerate bremmstrahlung loss, and $Q_{b, D}$ is the degenerate bremmstrahlung loss.~Equation \eqref{eq:loss} is implemented into MESA as an additional source of energy loss due to neutrinos.

The semi-Compton scattering loss rate is given as
\begin{equation}
    Q_{sC} \simeq 33 \alpha_{26} Y_e T_8^6 F_{\textrm{deg}} \frac{\textrm{erg}}{\textrm{g} \textrm{ s}},
    \label{eq:semiCompton}
\end{equation}
where  $Y_e = (Z/A)$ is the number of electrons per baryon, $T_8 = T / 10^8$K.~$F_{\textrm{deg}}$ encodes Pauli-blocking due to electron degeneracy and can be approximated as
\begin{equation}
    F_{\textrm{deg}} = \frac{1}{2}[1-\tanh f(\rho, T)]
\end{equation}
with
\begin{equation}
    f(\rho, T) = a \log_{10}\left[\frac{\rho}{\textrm{g} \textrm{ cm}^{-3}}\right] - b \log_{10} \left[\frac{T}{\textrm{K}}\right] + c
\end{equation}
where $a = 0.976$, $b = 0.1596$, and $c = 8.095$ per \cite{2021PDU....3200801C,2020arXiv200707889C}.~The bremsstrahlung loss rate  is broken into degenerate (D) and non-degenerate (ND) losses \cite{1995PhRvD..51.1495R}
\begin{equation}
    Q_{b, ND} \simeq 582 \alpha_{26} \frac{\textrm{erg}}{\textrm{g}\textrm{ s}} \rho_6 T^{5/2} F_{b, ND}
    \label{eq:ndBrem}
\end{equation}
and
\begin{equation}
    Q_{b, D} \simeq 10.8 \sum_{i = \textrm{ions}} \frac{X_i Z_i}{A_i} \alpha_{26} \frac{\textrm{erg}}{\textrm{g}\textrm{ s}} T_8^4 F_{b, D},
    \label{eq:dBrem}
\end{equation}
where $\rho_6 = \rho /(10^6$ g/cm$^3)$, $X_i$ is the mass fraction per ion, $Z_i$ is the number of electrons per ion, and $A_i$ is the number of proton and neutrons per ion,
\begin{equation}
    F_{b, D} = \frac{2}{3} \log(1 + 2\kappa^{-2}) + [(\kappa^2 + 2/5)\log(1+2\kappa^{-2})-2] \beta^2_F / 3,
\end{equation}
\begin{equation}
    F_{b, ND} = \sum_{i = \textrm{ions}} \frac{X_i Z_i^2}{A_i} \sum_{i = \textrm{ions}}  \frac{X_i Z_i }{A_i} + \frac{1}{\sqrt{2}}\left(\sum_{i = \textrm{ions}}  \frac{X_i Z_i }{A_i}\right)^2 ,
\end{equation} and
\begin{equation}
    \kappa^2 = \frac{k_s^2}{2 p_f^2} = \frac{4 \pi \alpha_{EM}}{T} \sum_{i=\textrm{ions}} n_{i} Z_i^2 +  n_{i} Z_i,
    \label{eq:kappa}
\end{equation} with 
$\beta_F = p_F / E_F$ where $p_F$ is the Fermi momentum and $E_F$ is the Fermi energy.~In \eqref{eq:kappa} above
\begin{equation}
    n_\textrm{ion} = \rho_{ion}\frac{X_{ion}}{A_{ion}} \mathcal{N}_A,
\end{equation} is the ion density of a particular ion.

\section{MCMC Results for Other Calibrations}
\label{sec:other_MCMC}

\begin{figure}
    \centering
    \includegraphics[width=\textwidth]{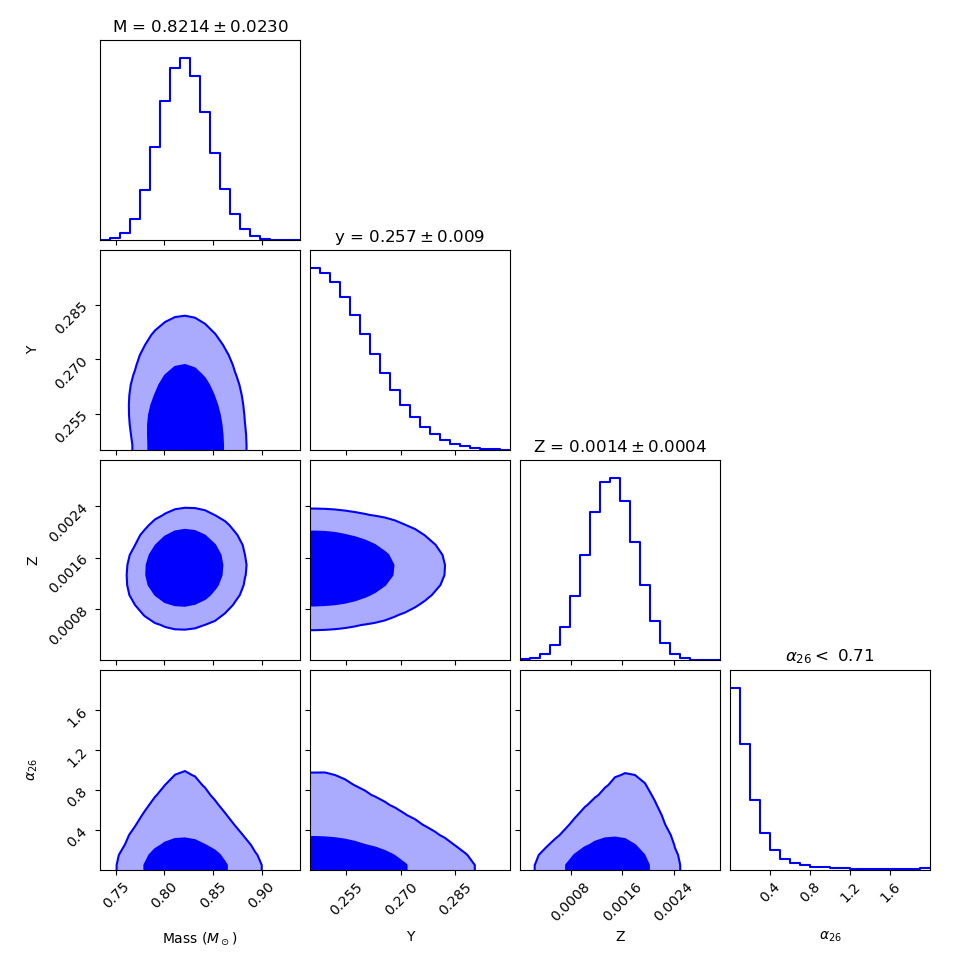}
    \caption{Same as \ref{fig:cornerTightPriorsStandard1} but for NGC 4258.}
    \label{fig:cornerTightPriorsStandard2}
\end{figure}

\begin{figure}
    \centering
    \includegraphics[width=\textwidth]{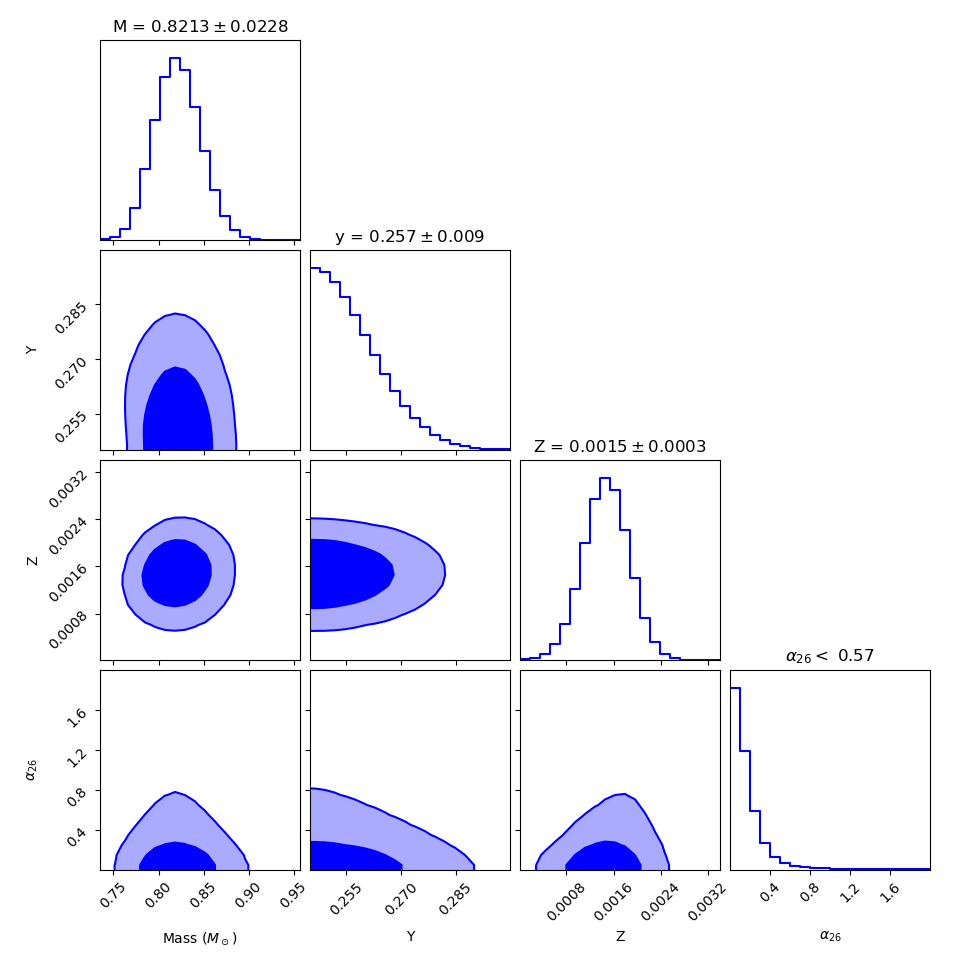}
    \caption{Same as \ref{fig:cornerTightPriorsStandard1} but for the LMC (Y19).}
    \label{fig:cornerTightPriorsStandard3}
\end{figure}

In this Appendix we show the results of our MCMC analysis for NGC 4258 (figures \ref{fig:cornerTightPriors2} and \ref{fig:cornerTightPriorsStandard2}) and the LMC (figures \ref{fig:cornerTightPriors3} and \ref{fig:cornerTightPriorsStandard3}).

\begin{figure}
    \centering
    \includegraphics[width=\textwidth]{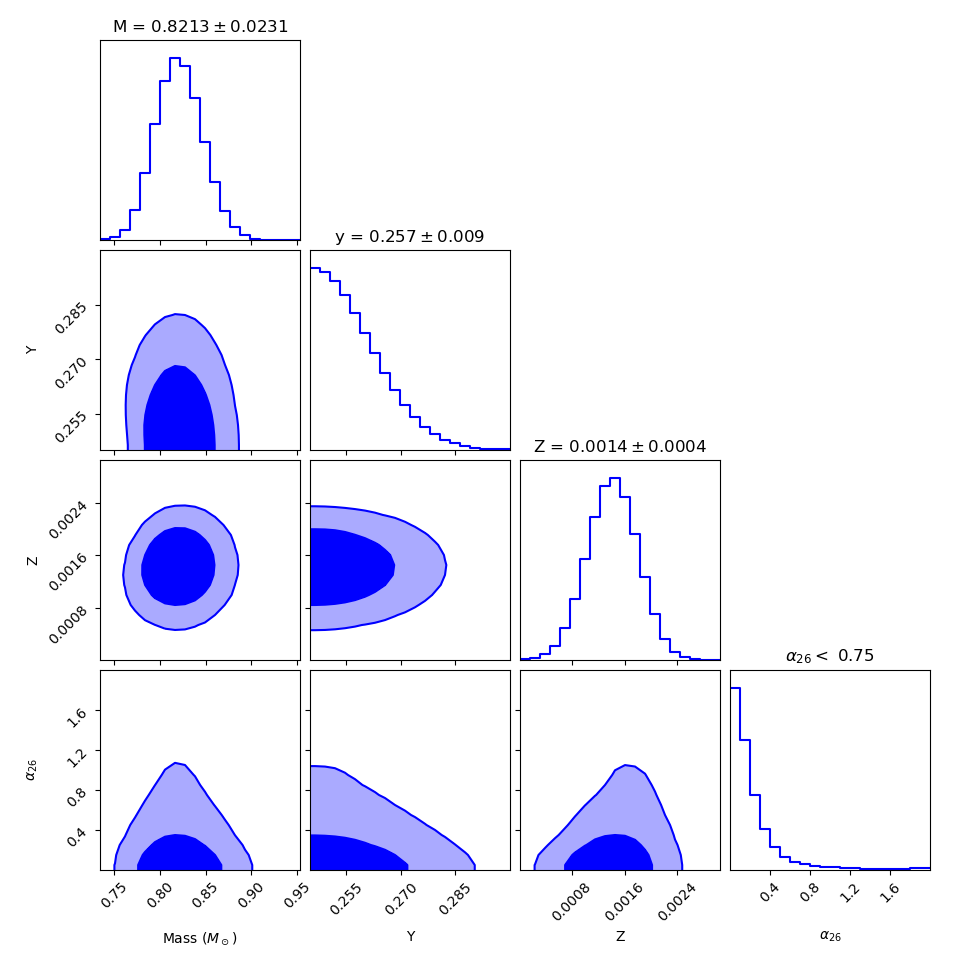}
    \caption{Same as \ref{fig:cornerTightPriorsStandard1} but for the LMC (F20).}
    \label{fig:cornerTightPriorsStandard4}
\end{figure}

\begin{figure}
    \centering
    \includegraphics[width=\textwidth]{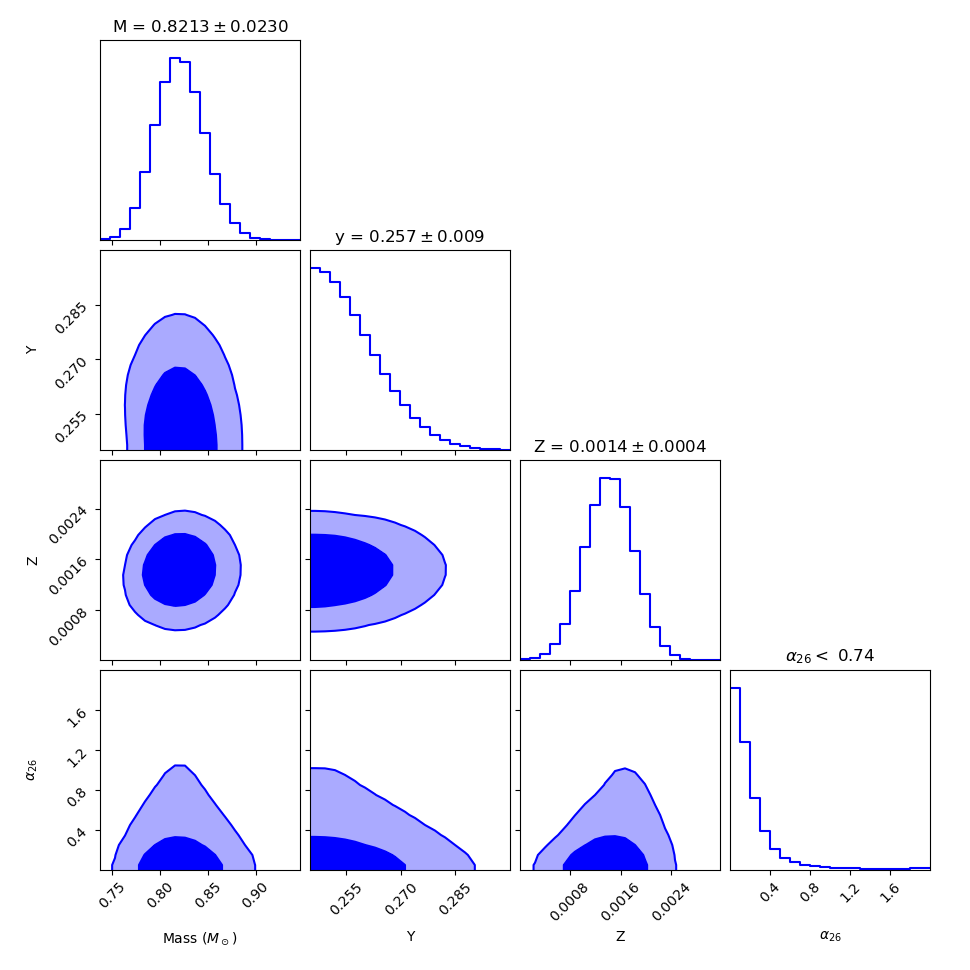}
    \caption{Same as \ref{fig:cornerTightPriorsStandard1} but with an updated distance from \cite{Baumgardt2021}.}
    \label{fig:cornerTightPriorsStandard5}
\end{figure}

\begin{figure}
    \centering
    \includegraphics[width=\textwidth]{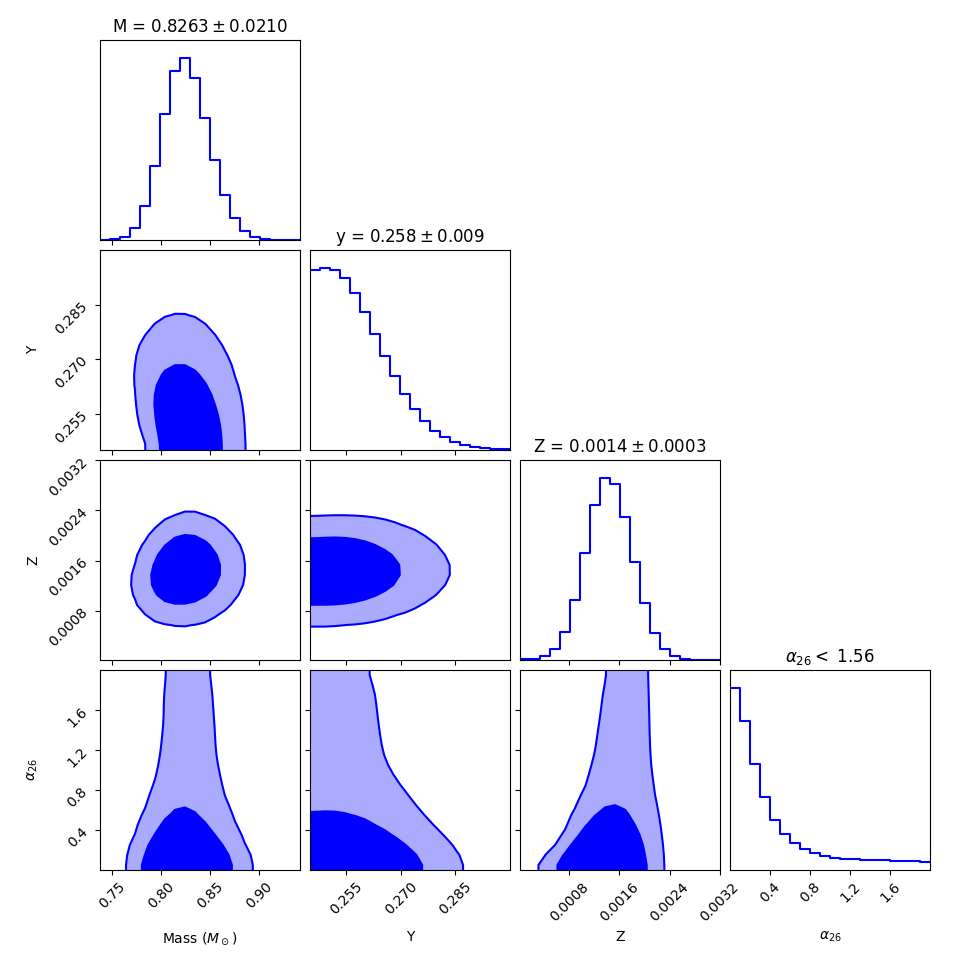}
    \caption{Same as \ref{fig:cornerTightPriors1} but for NGC 4258.}
    \label{fig:cornerTightPriors2}
\end{figure}

\begin{figure}
    \centering
    \includegraphics[width=\textwidth]{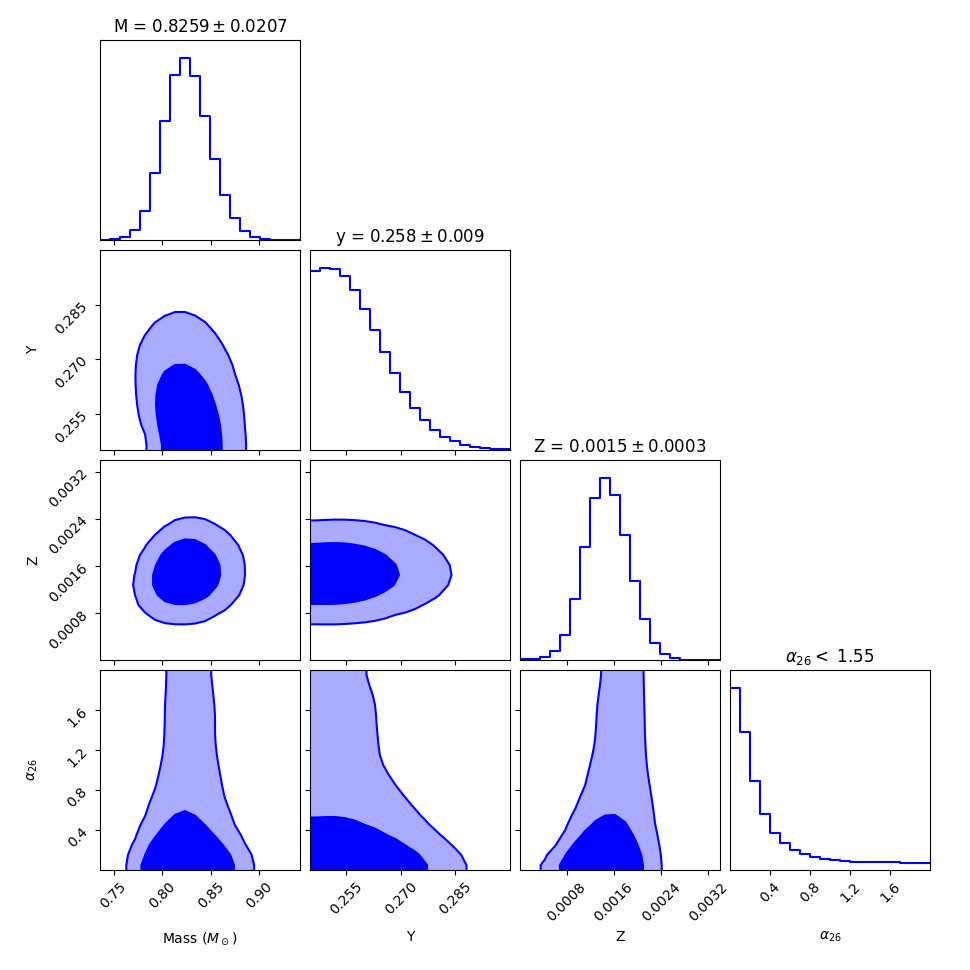}
    \caption{Same as \ref{fig:cornerTightPriors1} but for the LMC (Y19).}
    \label{fig:cornerTightPriors3}
\end{figure}

\begin{figure}
    \centering
    \includegraphics[width=\textwidth]{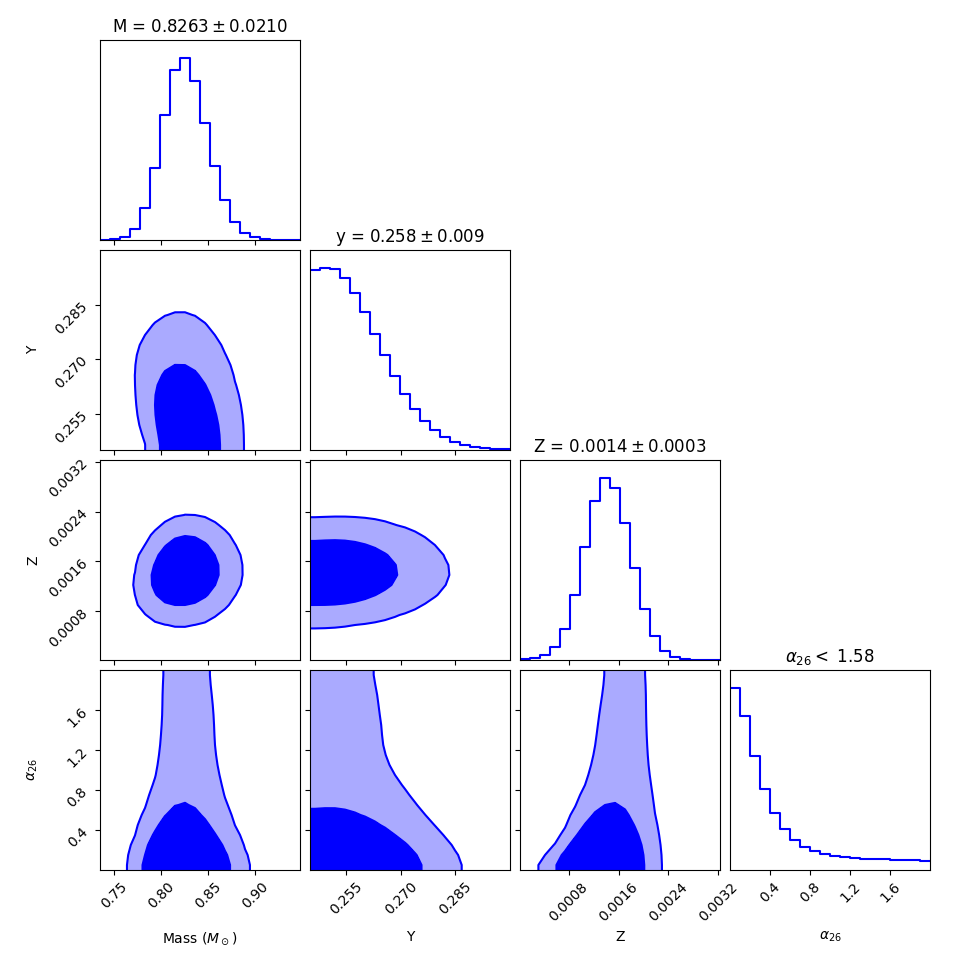}
    \caption{Same as \ref{fig:cornerTightPriors1} but for the LMC (F20).}
    \label{fig:cornerTightPriors4}
\end{figure}

\begin{figure}
    \centering
    \includegraphics[width=\textwidth]{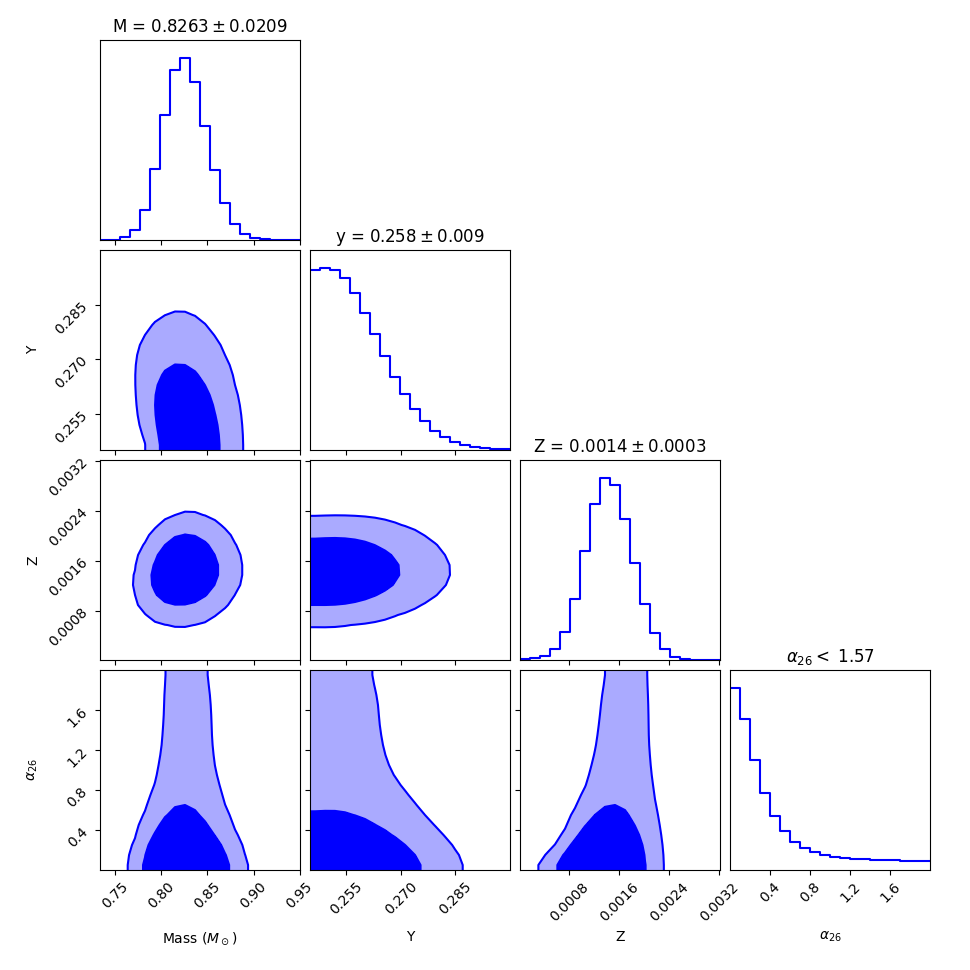}
    \caption{Same as \ref{fig:cornerTightPriors1} but with an updated distance from \cite{Baumgardt2021}.}
    \label{fig:cornerTightPriors5}
\end{figure}

\bibliographystyle{elsarticle-harv}
\bibliography{main.bib}
\end{document}

%% file: newcommands.tex

\newcommand{\tnm}[1]{{\tablenotemark{#1}}}
\newcommand{\tnt}[2]{{\tablenotetext{#1}{#2}}}
\newcommand{\fig}[2]{{fig.\,{#1}{#2}}}
\newcommand{\tab}[1]{{table\,{#1}}}
\newcommand{\eqn}[1]{{eq.\,{#1}}}

\newcommand{\Hawaii}{{Hawai`i}}

\newcommand{\eg}{{\it e.g.}}
\newcommand{\ie}{{\it i.e.}}
\newcommand{\etc}{{\it etc.}}
\newcommand{\etal}{{\it et~al.}}
\newcommand{\adhoc}{{\it ad~hoc}}
\newcommand{\insitu}{{\it in situ}}
\newcommand{\apriori}{{\it a~priori}}
\newcommand{\postfacto}{{\it post~facto}}

\newcommand{\half}{{\frac{1}{2}}}
\newcommand{\mean}[1]{\langle{#1}\rangle}
\newcommand{\dif}{\mathrm{d}}
\newcommand{\overbar}[1]{\mkern 1.5mu\overline{\mkern-1.5mu#1\mkern-1.5mu}\mkern 1.5mu}
\newcommand{\oforder}{\mathcal{O}}
\newcommand{\sci}[2]{{{#1}\times10^{#2}}}

\newcommand{\xbar}{\bar x}
\newcommand{\xyz}{(x,y,z)}
\newcommand{\xyzdot}{(\dot{x},\dot{y},\dot{z})}
\newcommand{\aei}{(a,e,i)}
\newcommand{\qei}{(q,e,i)}
\newcommand{\aeiH}{(a,e,i,H)}
\newcommand{\qeiD}{(q,e,i,D)}
\newcommand{\OoM}{(\Omega,\omega,M)}
\newcommand{\vo}{\vec{o}}
\newcommand{\vx}{\vec{x}}
\newcommand{\vxdot}{\vec{\dot x}}
\newcommand{\vxavg}{\mean{\vec{x}}}
\newcommand{\vy}{\vec{y}}
\newcommand{\vyavg}{\mean{\vec{y}}}
\newcommand{\vz}{\vec{z}}
\newcommand{\vzavg}{\mean{\vec{z}}}
\newcommand{\Havg}{\mean{H}}
\newcommand{\deltav}{\Delta v}
\newcommand{\node}{\Omega}
\newcommand{\aperi}{\omega}
\newcommand{\lperi}{\tilde\omega}

\newcommand{\todo}[2]{{\color{red}\bf #1 - #2}}
\newcommand{\XXX}{{\color{red}\bf XXX}}
\newcommand{\citepna}[1]{{{\color{red} (#1)}}}
\newcommand{\citetna}[1]{{{\color{red} #1}}}

\newcommand{\sn}{$S/N$}
\newcommand{\SN}{$S/N$}
\newcommand{\rd}{$^{rd}$}
\renewcommand{\st}{$^{st}$}
\newcommand{\nd}{$^{nd}$}
\newcommand{\fManx}{{\mathrm{f}_\mathrm{Manx}}}
\newcommand{\digesttwo}{{\texttt{digest2}}}

\newcommand{\HtwoO}{{H$_2$O}}
\newcommand{\CO}{{CO}}
\newcommand{\COtwo}{{CO$_2$}}

\newcommand{\arcdeg}{{^{\circ}}}
\newcommand{\arcmin}{^{\prime}}
\newcommand{\arcsec}{^{\prime\prime}}
\newcommand{\h}{^\mathrm{h}}
\newcommand{\m}{^\mathrm{m}}
\newcommand{\s}{^\mathrm{s}}
\newcommand{\Mpc}{\,\mathrm{Mpc}}
\newcommand{\kpc}{\,\mathrm{kpc}}
\newcommand{\pc}{\,\mathrm{pc}}
\newcommand{\au}{\,\mathrm{au}}
\newcommand{\km}{\,\mathrm{km}}
\newcommand{\kph}{\,\mathrm{km}/\mathrm{h}}
\newcommand{\kps}{\,\mathrm{km}\,\mathrm{s}^{-1}}
\newcommand{\ft}{\,\mathrm{ft}}
\newcommand{\meter}{\,\mathrm{m}}
\newcommand{\cm}{\,\mathrm{cm}}
\newcommand{\mm}{\,\mathrm{mm}}
\newcommand{\um}{\,\mu \mathrm{m}}
\newcommand{\nm}{\,\mathrm{nm}}
\newcommand{\rad}{\,\mathrm{rad}}
\newcommand{\rms}{\,\mathrm{(rms)}}
\newcommand{\anno}{\,\mathrm{a}}
\newcommand{\yr}{\,\mathrm{yr}}
\newcommand{\Myr}{\,\mathrm{Myr}}
\newcommand{\Gyr}{\,\mathrm{Gyr}}
\newcommand{\Day}{\,\mathrm{day}}
\newcommand{\days}{\,\mathrm{d}}
\newcommand{\dayperyear}{\,\mathrm{d}/\mathrm{yr}}
\newcommand{\vrk}{\,\mathrm{vrk}}
\newcommand{\degrees}{\,\mathrm{deg}}
\newcommand{\hours}{\,hours}
\newcommand{\hour}{\,\mathrm{h}}
\newcommand{\hourperday}{\,\mathrm{h}/\mathrm{d}}
\newcommand{\minute}{\,\mathrm{min}}
\newcommand{\second}{\,\mathrm{s}}
\newcommand{\mps}{\,\meter\,\second^{-1}}
\newcommand{\Hz}{\,\mathrm{Hz}}
\newcommand{\mags}{\,\mathrm{mag}}
\newcommand{\K}{\,\mathrm{K}}
\newcommand{\J}{\,\mathrm{J}}
\newcommand{\N}{\,\mathrm{N}}
\newcommand{\kg}{\,\mathrm{kg}}
\newcommand{\g}{\,\mathrm{g}}
\newcommand{\AMU}{\,\mathrm{AMU}}
\newcommand{\W}{\,\mathrm{W}}
\newcommand{\MW}{\,\mathrm{MW}}
\newcommand{\degC}{\arcdeg\mathrm{C}}
\newcommand{\degK}{\mathrm{K}}
\newcommand{\Jy}{\,\mathrm{Jy}}
\newcommand{\mJy}{\,\mathrm{mJy}}
\newcommand{\Mearth}{\,\mathrm{M}_\oplus}

\newcommand{\asteroid}[2]{{({#1})\,{#2}}}
\newcommand{\designation}[2]{{{#1}\,{#2}}}
\newcommand{\TC}{{2008\,TC$_3$}}
\newcommand{\RH}{{2006\,RH$_{120}$}}
\newcommand{\Bennu}{{(101955)\,Bennu}}

\newcommand{\Sthree}{{C/2014$\,$S3$\,$PANSTARRS}}
\newcommand{\Uone}{{1I/2017 U1 (‘Oumuamua)}}


\newcommand{\gps}{\ensuremath{g_{\rm P1}}}
\newcommand{\rps}{\ensuremath{r_{\rm P1}}}
\newcommand{\ips}{\ensuremath{i_{\rm P1}}}
\newcommand{\zps}{\ensuremath{z_{\rm P1}}}
\newcommand{\yps}{\ensuremath{y_{\rm P1}}}
\newcommand{\wps}{\ensuremath{w_{\rm P1}}}
\newcommand{\grizy}{\gps\rps\ips\zps\yps}
\newcommand{\JHK}{\ensuremath{JHK}}
\newcommand{\V}{\ensuremath{V}}

\newcommand{\PS}{\protect \hbox {Pan-STARRS}}
\newcommand{\PSone}{\protect \hbox {Pan-STARRS1}}
\newcommand{\PStwo}{\protect \hbox {Pan-STARRS2}}
\newcommand{\PSfour}{\protect \hbox {Pan-STARRS4}}
\newcommand{\knownserver}{{\tt known\_server}}


\newcommand\aj{AJ}
\newcommand\psj{{PSJ}}
\newcommand\araa{{ARA\&A}}
\newcommand\apj{{ApJ}}
\newcommand\apjl{{ApJL}}     
\newcommand\apjs{{ApJS}}
\newcommand\ao{{ApOpt}}
\newcommand\apss{{Ap\&SS}}
\newcommand\aap{{A\&A}}
\newcommand\aapr{{A\&A~Rv}}
\newcommand\aaps{{A\&AS}}
\newcommand\azh{{AZh}}
\newcommand\baas{{BAAS}}
\newcommand\icarus{{Icarus}}
\newcommand\jaavso{{JAAVSO}}  
\newcommand\jrasc{{JRASC}}
\newcommand\memras{{MmRAS}}
\newcommand\mnras{{MNRAS}}
\newcommand\pra{{PhRvA}}
\newcommand\prb{{PhRvB}}
\newcommand\prc{{PhRvC}}
\newcommand\prd{{PhRvD}}
\newcommand\pre{{PhRvE}}
\newcommand\prl{{PhRvL}}
\newcommand\pasp{{PASP}}
\newcommand\pasj{{PASJ}}
\newcommand\qjras{{QJRAS}}
\newcommand\skytel{{S\&T}}
\newcommand\solphys{{SoPh}}
\newcommand\sovast{{Soviet~Ast.}}
\newcommand\ssr{{SSRv}}
\newcommand\zap{{ZA}}
\newcommand\nat{{Nature}}
\newcommand\iaucirc{{IAUC}}
\newcommand\aplett{{Astrophys.~Lett.}}
\newcommand\apspr{{Astrophys.~Space~Phys.~Res.}}
\newcommand\bain{{BAN}}
\newcommand\fcp{{FCPh}}
\newcommand\gca{{GeoCoA}}
\newcommand\grl{{Geophys.~Res.~Lett.}}
\newcommand\jcp{{JChPh}}
\newcommand\jgr{{J.~Geophys.~Res.}}
\newcommand\jqsrt{{JQSRT}}
\newcommand\memsai{{MmSAI}}
\newcommand\nphysa{{NuPhA}}
\newcommand\physrep{{PhR}}
\newcommand\physscr{{PhyS}}
\newcommand\planss{{Planet.~Space~Sci.}}
\newcommand\procspie{{Proc.~SPIE}}

\newcommand\actaa{{AcA}}
\newcommand\caa{{ChA\&A}}
\newcommand\cjaa{{ChJA\&A}}
\newcommand\jcap{{JCAP}}
\newcommand\na{{NewA}}
\newcommand\nar{{NewAR}}
\newcommand\pasa{{PASA}}
\newcommand\rmxaa{{RMxAA}}

\newcommand\maps{{M\&PS}}
\newcommand\aas{{AAS Meeting Abstracts}}
\newcommand\dps{{AAS/DPS Meeting Abstracts}}

%% file: main.bib
@INPROCEEDINGS{2019SPIE11198E..13S,
       author = {{Sui}, Yu and {Zhang}, Xiaohui and {Huan}, Jiajia and {Hong}, Haifeng},
        title = "{Exploring data sampling techniques for imbalanced classification problems}",
    booktitle = {Fourth International Workshop on Pattern Recognition},
         year = 2019,
       editor = {{Jiang}, Xudong and {Chen}, Zhenxiang and {Chen}, Guojian},
       series = {Society of Photo-Optical Instrumentation Engineers (SPIE) Conference Series},
       volume = {11198},
        month = jul,
          eid = {1119813},
        pages = {1119813},
          doi = {10.1117/12.2540457},
       adsurl = {https://ui.adsabs.harvard.edu/abs/2019SPIE11198E..13S}
}

@ARTICLE{1998SSRv...85..161G,
       author = {{Grevesse}, N. and {Sauval}, A.~J.},
        title = "{Standard Solar Composition}",
      journal = {\ssr},
         year = 1998,
        month = may,
       volume = {85},
        pages = {161-174},
          doi = {10.1023/A:1005161325181},
       adsurl = {https://ui.adsabs.harvard.edu/abs/1998SSRv...85..161G}
}

@ARTICLE{2019MNRAS.482.1352M,
       author = {{McClintock}, T. and {Varga}, T.~N. and {Gruen}, D. and {Rozo}, E. and {Rykoff}, E.~S. and {Shin}, T. and {Melchior}, P. and {DeRose}, J. and {Seitz}, S. and {Dietrich}, J.~P. and {Sheldon}, E. and {Zhang}, Y. and {von der Linden}, A. and {Jeltema}, T. and {Mantz}, A.~B. and {Romer}, A.~K. and {Allen}, S. and {Becker}, M.~R. and {Bermeo}, A. and {Bhargava}, S. and {Costanzi}, M. and {Everett}, S. and {Farahi}, A. and {Hamaus}, N. and {Hartley}, W.~G. and {Hollowood}, D.~L. and {Hoyle}, B. and {Israel}, H. and {Li}, P. and {MacCrann}, N. and {Morris}, G. and {Palmese}, A. and {Plazas}, A.~A. and {Pollina}, G. and {Rau}, M.~M. and {Simet}, M. and {Soares-Santos}, M. and {Troxel}, M.~A. and {Vergara Cervantes}, C. and {Wechsler}, R.~H. and {Zuntz}, J. and {Abbott}, T.~M.~C. and {Abdalla}, F.~B. and {Allam}, S. and {Annis}, J. and {Avila}, S. and {Bridle}, S.~L. and {Brooks}, D. and {Burke}, D.~L. and {Carnero Rosell}, A. and {Carrasco Kind}, M. and {Carretero}, J. and {Castander}, F.~J. and {Crocce}, M. and {Cunha}, C.~E. and {D'Andrea}, C.~B. and {da Costa}, L.~N. and {Davis}, C. and {De Vicente}, J. and {Diehl}, H.~T. and {Doel}, P. and {Drlica-Wagner}, A. and {Evrard}, A.~E. and {Flaugher}, B. and {Fosalba}, P. and {Frieman}, J. and {Garc{\'\i}a-Bellido}, J. and {Gaztanaga}, E. and {Gerdes}, D.~W. and {Giannantonio}, T. and {Gruendl}, R.~A. and {Gutierrez}, G. and {Honscheid}, K. and {James}, D.~J. and {Kirk}, D. and {Krause}, E. and {Kuehn}, K. and {Lahav}, O. and {Li}, T.~S. and {Lima}, M. and {March}, M. and {Marshall}, J.~L. and {Menanteau}, F. and {Miquel}, R. and {Mohr}, J.~J. and {Nord}, B. and {Ogando}, R.~L.~C. and {Roodman}, A. and {Sanchez}, E. and {Scarpine}, V. and {Schindler}, R. and {Sevilla-Noarbe}, I. and {Smith}, M. and {Smith}, R.~C. and {Sobreira}, F. and {Suchyta}, E. and {Swanson}, M.~E.~C. and {Tarle}, G. and {Tucker}, D.~L. and {Vikram}, V. and {Walker}, A.~R. and {Weller}, J. and {DES Collaboration}},
        title = "{Dark Energy Survey Year 1 results: weak lensing mass calibration of redMaPPer galaxy clusters}",
      journal = {\mnras},
         year = 2019,
        month = jan,
       volume = {482},
       number = {1},
        pages = {1352-1378},
          doi = {10.1093/mnras/sty2711},
archivePrefix = {arXiv},
       eprint = {1805.00039},
 primaryClass = {astro-ph.CO},
       adsurl = {https://ui.adsabs.harvard.edu/abs/2019MNRAS.482.1352M}
}

@ARTICLE{Viaux2013,
       author = {{Viaux}, N. and {Catelan}, M. and {Stetson}, P.~B. and {Raffelt}, G.~G. and {Redondo}, J. and {Valcarce}, A.~A.~R. and {Weiss}, A.},
        title = "{Particle-physics constraints from the globular cluster M5: neutrino dipole moments}",
      journal = {\aap},
         year = 2013,
        month = oct,
       volume = {558},
          eid = {A12},
        pages = {A12},
          doi = {10.1051/0004-6361/201322004},
archivePrefix = {arXiv},
       eprint = {1308.4627},
 primaryClass = {astro-ph.SR},
       adsurl = {https://ui.adsabs.harvard.edu/abs/2013A&A...558A..12V}
}

@misc{chollet2015keras,
  title={Keras},
  author={Chollet, Fran\c{c}ois and others},
  year={2015},
  howpublished={\url{https://keras.io}},
}

@ARTICLE{2014arXiv1412.6980K,
       author = {{Kingma}, Diederik P. and {Ba}, Jimmy},
        title = "{Adam: A Method for Stochastic Optimization}",
      journal = {arXiv e-prints},
         year = 2014,
        month = dec,
          eid = {arXiv:1412.6980},
        pages = {arXiv:1412.6980},
archivePrefix = {arXiv},
       eprint = {1412.6980},
 primaryClass = {cs.LG},
       adsurl = {https://ui.adsabs.harvard.edu/abs/2014arXiv1412.6980K}
}

@misc{tensorflow2015-whitepaper,
title={ {TensorFlow}: Large-Scale Machine Learning on Heterogeneous Systems},
url={https://www.tensorflow.org/},
note={Software available from tensorflow.org},
author={
    Mart\'{i}n~Abadi and
    Ashish~Agarwal and
    Paul~Barham and
    Eugene~Brevdo and
    Zhifeng~Chen and
    Craig~Citro and
    Greg~S.~Corrado and
    Andy~Davis and
    Jeffrey~Dean and
    Matthieu~Devin and
    Sanjay~Ghemawat and
    Ian~Goodfellow and
    Andrew~Harp and
    Geoffrey~Irving and
    Michael~Isard and
    Yangqing Jia and
    Rafal~Jozefowicz and
    Lukasz~Kaiser and
    Manjunath~Kudlur and
    Josh~Levenberg and
    Dandelion~Man\'{e} and
    Rajat~Monga and
    Sherry~Moore and
    Derek~Murray and
    Chris~Olah and
    Mike~Schuster and
    Jonathon~Shlens and
    Benoit~Steiner and
    Ilya~Sutskever and
    Kunal~Talwar and
    Paul~Tucker and
    Vincent~Vanhoucke and
    Vijay~Vasudevan and
    Fernanda~Vi\'{e}gas and
    Oriol~Vinyals and
    Pete~Warden and
    Martin~Wattenberg and
    Martin~Wicke and
    Yuan~Yu and
    Xiaoqiang~Zheng},
  year={2015},
}

@ARTICLE{2020PhRvD.102b3007D,
       author = {{Desmond}, Harry and {Sakstein}, Jeremy},
        title = "{Screened fifth forces lower the TRGB-calibrated Hubble constant too}",
      journal = {\prd},
         year = 2020,
        month = jul,
       volume = {102},
       number = {2},
          eid = {023007},
        pages = {023007},
          doi = {10.1103/PhysRevD.102.023007},
archivePrefix = {arXiv},
       eprint = {2003.12876},
 primaryClass = {astro-ph.CO},
       adsurl = {https://ui.adsabs.harvard.edu/abs/2020PhRvD.102b3007D}
}

@INCOLLECTION{Sokal,
        author = {{Sokal}, Alan},
        title = "{Monte Carlo Methods in Statisitical Mechanics: Foundations and New Algorithms}",
        booktitle = {Functional Integration: Basics and Applications},
        year = {1997},
        editor = {{DeWitt-Morette}, Cecile and {Cartier}, Pierre and {Folacci}, Antoine},
        series = {NATO ASI Series},
        pages = {131-192},
        publisher = {Springer},
        address = {Boston, MA, USA}
}

@BOOK{1996slfp.book.....R,
       author = {{Raffelt}, Georg G.},
        title = "{Stars as laboratories for fundamental physics : the astrophysics of neutrinos, axions, and other weakly interacting particles}",
         year = 1996,
       adsurl = {https://ui.adsabs.harvard.edu/abs/1996slfp.book.....R}
}

@ARTICLE{2013PASP..125..306F,
       author = {{Foreman-Mackey}, Daniel and {Hogg}, David W. and {Lang}, Dustin and {Goodman}, Jonathan},
        title = "{emcee: The MCMC Hammer}",
      journal = {\pasp},
         year = 2013,
        month = mar,
       volume = {125},
       number = {925},
        pages = {306},
          doi = {10.1086/670067},
archivePrefix = {arXiv},
       eprint = {1202.3665},
 primaryClass = {astro-ph.IM},
       adsurl = {https://ui.adsabs.harvard.edu/abs/2013PASP..125..306F}
}

@ARTICLE{2001ApJ...556..635B,
       author = {{Bellazzini}, Michele and {Ferraro}, Francesco R. and {Pancino}, Elena},
        title = "{A Step toward the Calibration of the Red Giant Branch Tip as a Standard Candle}",
      journal = {\apj},
         year = 2001,
        month = aug,
       volume = {556},
       number = {2},
        pages = {635-640},
          doi = {10.1086/321613},
archivePrefix = {arXiv},
       eprint = {astro-ph/0104114},
 primaryClass = {astro-ph},
       adsurl = {https://ui.adsabs.harvard.edu/abs/2001ApJ...556..635B}
}

@BOOK{2004sipp.book.....H,
       author = {{Hansen}, Carl J. and {Kawaler}, Steven D. and {Trimble}, Virginia},
        title = "{Stellar interiors : physical principles, structure, and evolution}",
         year = 2004,
       adsurl = {https://ui.adsabs.harvard.edu/abs/2004sipp.book.....H}
}

@BOOK{2013sse..book.....K,
       author = {{Kippenhahn}, Rudolf and {Weigert}, Alfred and {Weiss}, Achim},
        title = "{Stellar Structure and Evolution}",
         year = 2013,
          doi = {10.1007/978-3-642-30304-3},
       adsurl = {https://ui.adsabs.harvard.edu/abs/2013sse..book.....K}
}

@ARTICLE{2017ApJ...835...28J,
       author = {{Jang}, In Sung and {Lee}, Myung Gyoon},
        title = "{The Tip of the Red Giant Branch Distances to Type Ia Supernova Host Galaxies. IV. Color Dependence and Zero-point Calibration}",
      journal = {\apj},
         year = 2017,
        month = jan,
       volume = {835},
       number = {1},
          eid = {28},
        pages = {28},
          doi = {10.3847/1538-4357/835/1/28},
archivePrefix = {arXiv},
       eprint = {1611.05040},
 primaryClass = {astro-ph.GA},
       adsurl = {https://ui.adsabs.harvard.edu/abs/2017ApJ...835...28J}
}

@ARTICLE{2019ApJ...886...61Y,
       author = {{Yuan}, Wenlong and {Riess}, Adam G. and {Macri}, Lucas M. and {Casertano}, Stefano and {Scolnic}, Daniel M.},
        title = "{Consistent Calibration of the Tip of the Red Giant Branch in the Large Magellanic Cloud on the Hubble Space Telescope Photometric System and a Redetermination of the Hubble Constant}",
      journal = {\apj},
         year = 2019,
        month = nov,
       volume = {886},
       number = {1},
          eid = {61},
        pages = {61},
          doi = {10.3847/1538-4357/ab4bc9},
archivePrefix = {arXiv},
       eprint = {1908.00993},
 primaryClass = {astro-ph.GA},
       adsurl = {https://ui.adsabs.harvard.edu/abs/2019ApJ...886...61Y}
}

@ARTICLE{Ferraro2006,
       author = {{Ferraro}, Francesco R. and {Valenti}, Elena and {Straniero}, Oscar and {Origlia}, Livia},
        title = "{An Empirical Calibration of the Mixing-Length Parameter {\ensuremath{\alpha}}}",
      journal = {\apj},
         year = 2006,
        month = may,
       volume = {642},
       number = {1},
        pages = {225-229},
          doi = {10.1086/500803},
archivePrefix = {arXiv},
       eprint = {astro-ph/0601159},
 primaryClass = {astro-ph},
       adsurl = {https://ui.adsabs.harvard.edu/abs/2006ApJ...642..225F}
}

@ARTICLE{Salaris1996,
       author = {{Salaris}, M. and {Cassisi}, S.},
        title = "{New molecular opacities and effective temperature of RGB stellar models}",
      journal = {\aap},
         year = 1996,
        month = jan,
       volume = {305},
        pages = {858},
       adsurl = {https://ui.adsabs.harvard.edu/abs/1996A&A...305..858S}
}

@ARTICLE{2020ApJ...891...57F,
       author = {{Freedman}, Wendy L. and {Madore}, Barry F. and {Hoyt}, Taylor and {Jang}, In Sung and {Beaton}, Rachael and {Lee}, Myung Gyoon and {Monson}, Andrew and {Neeley}, Jill and {Rich}, Jeffrey},
        title = "{Calibration of the Tip of the Red Giant Branch}",
      journal = {\apj},
         year = 2020,
        month = mar,
       volume = {891},
       number = {1},
          eid = {57},
        pages = {57},
          doi = {10.3847/1538-4357/ab7339},
archivePrefix = {arXiv},
       eprint = {2002.01550},
 primaryClass = {astro-ph.GA},
       adsurl = {https://ui.adsabs.harvard.edu/abs/2020ApJ...891...57F}
}

@ARTICLE{2010CAMCS...5...65G,
       author = {{Goodman}, Jonathan and {Weare}, Jonathan},
        title = "{Ensemble samplers with affine invariance}",
      journal = {Communications in Applied Mathematics and Computational Science},
         year = 2010,
        month = jan,
       volume = {5},
       number = {1},
        pages = {65-80},
          doi = {10.2140/camcos.2010.5.65},
       adsurl = {https://ui.adsabs.harvard.edu/abs/2010CAMCS...5...65G}
}

@ARTICLE{2010ApJ...719..865S,
       author = {{Serenelli}, Aldo M. and {Basu}, Sarbani},
        title = "{Determining the Initial Helium Abundance of the Sun}",
      journal = {\apj},
         year = 2010,
        month = aug,
       volume = {719},
       number = {1},
        pages = {865-872},
          doi = {10.1088/0004-637X/719/1/865},
archivePrefix = {arXiv},
       eprint = {1006.0244},
 primaryClass = {astro-ph.SR},
       adsurl = {https://ui.adsabs.harvard.edu/abs/2010ApJ...719..865S}
}

@ARTICLE{2013PhRvL.111w1301V,
       author = {{Viaux}, N. and {Catelan}, M. and {Stetson}, P.~B. and {Raffelt}, G.~G. and {Redondo}, J. and {Valcarce}, A.~A.~R. and {Weiss}, A.},
        title = "{Neutrino and Axion Bounds from the Globular Cluster M5 (NGC 5904)}",
      journal = {\prl},
         year = 2013,
        month = dec,
       volume = {111},
       number = {23},
          eid = {231301},
        pages = {231301},
          doi = {10.1103/PhysRevLett.111.231301},
archivePrefix = {arXiv},
       eprint = {1311.1669},
 primaryClass = {astro-ph.SR},
       adsurl = {https://ui.adsabs.harvard.edu/abs/2013PhRvL.111w1301V}
}

@ARTICLE{Serenelli2017,
       author = {{Serenelli}, A. and {Weiss}, A. and {Cassisi}, S. and {Salaris}, M. and {Pietrinferni}, A.},
        title = "{The brightness of the red giant branch tip. Theoretical framework, a set of reference models, and predicted observables}",
      journal = {\aap},
         year = 2017,
        month = oct,
       volume = {606},
          eid = {A33},
        pages = {A33},
          doi = {10.1051/0004-6361/201731004},
archivePrefix = {arXiv},
       eprint = {1706.09910},
 primaryClass = {astro-ph.SR},
       adsurl = {https://ui.adsabs.harvard.edu/abs/2017A&A...606A..33S}
}

@ARTICLE{2019ApJ...882...34F,
       author = {{Freedman}, Wendy L. and {Madore}, Barry F. and {Hatt}, Dylan and {Hoyt}, Taylor J. and {Jang}, In Sung and {Beaton}, Rachael L. and {Burns}, Christopher R. and {Lee}, Myung Gyoon and {Monson}, Andrew J. and {Neeley}, Jillian R. and {Phillips}, M.~M. and {Rich}, Jeffrey A. and {Seibert}, Mark},
        title = "{The Carnegie-Chicago Hubble Program. VIII. An Independent Determination of the Hubble Constant Based on the Tip of the Red Giant Branch}",
      journal = {\apj},
         year = 2019,
        month = sep,
       volume = {882},
       number = {1},
          eid = {34},
        pages = {34},
          doi = {10.3847/1538-4357/ab2f73},
archivePrefix = {arXiv},
       eprint = {1907.05922},
 primaryClass = {astro-ph.CO},
       adsurl = {https://ui.adsabs.harvard.edu/abs/2019ApJ...882...34F}
}

@ARTICLE{corner,
      doi = {10.21105/joss.00024},
      url = {https://doi.org/10.21105/joss.00024},
      year  = {2016},
      month = {jun},
      publisher = {The Open Journal},
      volume = {1},
      number = {2},
      pages = {24},
      author = {Daniel Foreman-Mackey},
      title = {corner.py: Scatterplot matrices in Python},
      journal = {The Journal of Open Source Software}
}

@ARTICLE{2008ApJS..178...89D,
       author = {{Dotter}, Aaron and {Chaboyer}, Brian and {Jevremovi{\'c}}, Darko and {Kostov}, Veselin and {Baron}, E. and {Ferguson}, Jason W.},
        title = "{The Dartmouth Stellar Evolution Database}",
      journal = {\apjs},
         year = 2008,
        month = sep,
       volume = {178},
       number = {1},
        pages = {89-101},
          doi = {10.1086/589654},
archivePrefix = {arXiv},
       eprint = {0804.4473},
 primaryClass = {astro-ph},
       adsurl = {https://ui.adsabs.harvard.edu/abs/2008ApJS..178...89D}
}

@ARTICLE{2008A&A...486..951G,
       author = {{Gustafsson}, B. and {Edvardsson}, B. and {Eriksson}, K. and {J{\o}rgensen}, U.~G. and {Nordlund}, {\r{A}}. and {Plez}, B.},
        title = "{A grid of MARCS model atmospheres for late-type stars. I. Methods and general properties}",
      journal = {\aap},
         year = 2008,
        month = aug,
       volume = {486},
       number = {3},
        pages = {951-970},
          doi = {10.1051/0004-6361:200809724},
archivePrefix = {arXiv},
       eprint = {0805.0554},
 primaryClass = {astro-ph},
       adsurl = {https://ui.adsabs.harvard.edu/abs/2008A&A...486..951G}
}

@ARTICLE{2021A&A...651A.101L,
       author = {{Lopes}, Jos{\'e} and {Lopes}, Il{\'\i}dio},
        title = "{Dark matter capture and annihilation in stars: Impact on the red giant branch tip}",
      journal = {\aap},
         year = 2021,
        month = jul,
       volume = {651},
          eid = {A101},
        pages = {A101},
          doi = {10.1051/0004-6361/202140750},
archivePrefix = {arXiv},
       eprint = {2107.13885},
 primaryClass = {astro-ph.SR},
       adsurl = {https://ui.adsabs.harvard.edu/abs/2021A&A...651A.101L}
}

@ARTICLE{2021PDU....3200801C,
       author = {{Croon}, Djuna and {McDermott}, Samuel D. and {Sakstein}, Jeremy},
        title = "{Missing in axion: Where are XENON1T's big black holes?}",
      journal = {Physics of the Dark Universe},
         year = 2021,
        month = may,
       volume = {32},
          eid = {100801},
        pages = {100801},
          doi = {10.1016/j.dark.2021.100801},
archivePrefix = {arXiv},
       eprint = {2007.00650},
 primaryClass = {hep-ph},
       adsurl = {https://ui.adsabs.harvard.edu/abs/2021PDU....3200801C}
}

@ARTICLE{2020arXiv200707889C,
       author = {{Croon}, Djuna and {McDermott}, Samuel D. and {Sakstein}, Jeremy},
        title = "{Missing in Action: New Physics and the Black Hole Mass Gap}",
      journal = {\prd},
         year = 2020,
        month = jul,
          eid = {arXiv:2007.07889},
        pages = {arXiv:2007.07889},
archivePrefix = {arXiv},
       eprint = {2007.07889},
 primaryClass = {gr-qc},
       adsurl = {https://ui.adsabs.harvard.edu/abs/2020arXiv200707889C}
}

@ARTICLE{1995PhRvD..51.1495R,
       author = {{Raffelt}, Georg and {Weiss}, Achim},
        title = "{Red giant bound on the axion-electron coupling reexamined}",
      journal = {\prd},
         year = 1995,
        month = feb,
       volume = {51},
       number = {4},
        pages = {1495-1498},
          doi = {10.1103/PhysRevD.51.1495},
archivePrefix = {arXiv},
       eprint = {hep-ph/9410205},
 primaryClass = {hep-ph},
       adsurl = {https://ui.adsabs.harvard.edu/abs/1995PhRvD..51.1495R}
}

@inproceedings{Ayala:2015juy,
    author = "Ayala, Adrian and Straniero, Oscar and Giannotti, Maurizio and Mirizzi, Alessandro and Dominguez, Inma",
    title = "{Effects of Hidden Photons during the Red Giant Branch (RGB) Phase}",
    booktitle = "{11th Patras Workshop on Axions, WIMPs and WISPs}",
    doi = "10.3204/DESY-PROC-2015-02/ayala_adrian",
    pages = "189--192",
    year = "2015"
}

@ARTICLE{2015APh....70....1A,
       author = {{Arceo-D{\'\i}az}, S. and {Schr{\"o}der}, K. -P. and {Zuber}, K. and {Jack}, D.},
        title = "{Constraint on the magnetic dipole moment of neutrinos by the tip-RGB luminosity in {\ensuremath{\omega}}-Centauri}",
      journal = {Astroparticle Physics},
         year = 2015,
        month = oct,
       volume = {70},
        pages = {1-11},
          doi = {10.1016/j.astropartphys.2015.03.006},
       adsurl = {https://ui.adsabs.harvard.edu/abs/2015APh....70....1A}
}

@ARTICLE{Baumgardt2019,
       author = {{Baumgardt}, H. and {Hilker}, M. and {Sollima}, A. and {Bellini}, A.},
        title = "{Mean proper motions, space orbits, and velocity dispersion profiles of Galactic globular clusters derived from Gaia DR2 data}",
      journal = {\mnras},
         year = 2019,
        month = feb,
       volume = {482},
       number = {4},
        pages = {5138-5155},
          doi = {10.1093/mnras/sty2997},
archivePrefix = {arXiv},
       eprint = {1811.01507},
 primaryClass = {astro-ph.GA},
       adsurl = {https://ui.adsabs.harvard.edu/abs/2019MNRAS.482.5138B}
}

@ARTICLE{Baumgardt2021,
       author = {{Baumgardt}, H. and {Vasiliev}, E.},
        title = "{Accurate distances to Galactic globular clusters through a combination of Gaia EDR3, HST, and literature data}",
      journal = {\mnras},
         year = 2021,
        month = aug,
       volume = {505},
       number = {4},
        pages = {5957-5977},
          doi = {10.1093/mnras/stab1474},
archivePrefix = {arXiv},
       eprint = {2105.09526},
 primaryClass = {astro-ph.GA},
       adsurl = {https://ui.adsabs.harvard.edu/abs/2021MNRAS.505.5957B}
}

@ARTICLE{Straniero2020,
       author = {{Straniero}, O. and {Pallanca}, C. and {Dalessandro}, E. and {Dom{\'\i}nguez}, I. and {Ferraro}, F.~R. and {Giannotti}, M. and {Mirizzi}, A. and {Piersanti}, L.},
        title = "{The RGB tip of galactic globular clusters and the revision of the axion-electron coupling bound}",
      journal = {\aap},
         year = 2020,
        month = dec,
       volume = {644},
          eid = {A166},
        pages = {A166},
          doi = {10.1051/0004-6361/202038775},
archivePrefix = {arXiv},
       eprint = {2010.03833},
 primaryClass = {astro-ph.SR},
       adsurl = {https://ui.adsabs.harvard.edu/abs/2020A&A...644A.166S}
}

@ARTICLE{2020JCAP...11..029A,
       author = {{Alonso-{\'A}lvarez}, Gonzalo and {Ertas}, Fatih and {Jaeckel}, Joerg and {Kahlhoefer}, Felix and {Thormaehlen}, Lennert J.},
        title = "{Hidden photon dark matter in the light of XENON1T and stellar cooling}",
      journal = {\jcap},
         year = 2020,
        month = nov,
       volume = {2020},
       number = {11},
          eid = {029},
        pages = {029},
          doi = {10.1088/1475-7516/2020/11/029},
archivePrefix = {arXiv},
       eprint = {2006.11243},
 primaryClass = {hep-ph},
       adsurl = {https://ui.adsabs.harvard.edu/abs/2020JCAP...11..029A}
}

@ARTICLE{2007CSE.....9...90H,
       author = {{Hunter}, John D.},
        title = "{Matplotlib: A 2D Graphics Environment}",
      journal = {Computing in Science and Engineering},
         year = 2007,
        month = may,
       volume = {9},
       number = {3},
        pages = {90-95},
          doi = {10.1109/MCSE.2007.55},
       adsurl = {https://ui.adsabs.harvard.edu/abs/2007CSE.....9...90H}
}

@software{pandasSoftware,
    author       = {The pandas development team},
    title        = {pandas-dev/pandas: Pandas},
    month        = feb,
    year         = 2020,
    publisher    = {Zenodo},
    version      = {latest},
    doi          = {10.5281/zenodo.3509134},
    url          = {https://doi.org/10.5281/zenodo.3509134}
}

@InProceedings{pandasDataStructure,
  author    = { {W}es {M}c{K}inney },
  title     = { {D}ata {S}tructures for {S}tatistical {C}omputing in {P}ython },
  booktitle = { {P}roceedings of the 9th {P}ython in {S}cience {C}onference },
  pages     = { 56 - 61 },
  year      = { 2010 },
  editor    = { {S}t\'efan van der {W}alt and {J}arrod {M}illman },
  doi       = { 10.25080/Majora-92bf1922-00a }
}

@ARTICLE{2020Natur.585..357H,
       author = {{Harris}, Charles R. and {Millman}, K. Jarrod and {van der Walt}, St{\'e}fan J. and {Gommers}, Ralf and {Virtanen}, Pauli and {Cournapeau}, David and {Wieser}, Eric and {Taylor}, Julian and {Berg}, Sebastian and {Smith}, Nathaniel J. and {Kern}, Robert and {Picus}, Matti and {Hoyer}, Stephan and {van Kerkwijk}, Marten H. and {Brett}, Matthew and {Haldane}, Allan and {del R{\'\i}o}, Jaime Fern{\'a}ndez and {Wiebe}, Mark and {Peterson}, Pearu and {G{\'e}rard-Marchant}, Pierre and {Sheppard}, Kevin and {Reddy}, Tyler and {Weckesser}, Warren and {Abbasi}, Hameer and {Gohlke}, Christoph and {Oliphant}, Travis E.},
        title = "{Array programming with NumPy}",
      journal = {\nat},
         year = 2020,
        month = sep,
       volume = {585},
       number = {7825},
        pages = {357-362},
          doi = {10.1038/s41586-020-2649-2},
archivePrefix = {arXiv},
       eprint = {2006.10256},
 primaryClass = {cs.MS},
       adsurl = {https://ui.adsabs.harvard.edu/abs/2020Natur.585..357H}
}

@ARTICLE{Paxton2011,
  author = {{Paxton}, B. and {Bildsten}, L. and {Dotter}, A. and {Herwig}, F. and {Lesaffre}, P. and {Timmes}, F.},
  title = {{Modules for Experiments in Stellar Astrophysics (MESA)}},
  journal = {\apjs},
  archivePrefix = {arXiv},
  eprint = {1009.1622},
  primaryClass = {astro-ph.SR},
  year = {2011},
  month = {jan},
  volume = {192},
  eid = {3},
  pages = {3},
  doi = {10.1088/0067-0049/192/1/3},
  adsurl = {https://ui.adsabs.harvard.edu/abs/2011ApJS..192....3P},
}

@ARTICLE{Paxton2013,
  author = {{Paxton}, B. and {Cantiello}, M. and {Arras}, P. and {Bildsten}, L. and {Brown}, E.~F. and {Dotter}, A. and {Mankovich}, C. and {Montgomery}, M.~H. and {Stello}, D. and {Timmes}, F.~X. and {Townsend}, R.},
  title = {{Modules for Experiments in Stellar Astrophysics (MESA): Planets, Oscillations, Rotation, and Massive Stars}},
  journal = {\apjs},
  archivePrefix = {arXiv},
  eprint = {1301.0319},
  primaryClass = {astro-ph.SR},
  year = {2013},
  month = {sep},
  volume = {208},
  eid = {4},
  pages = {4},
  doi = {10.1088/0067-0049/208/1/4},
  adsurl = {https://ui.adsabs.harvard.edu/abs/2013ApJS..208....4P},
}

@ARTICLE{Paxton2015,
  author = {{Paxton}, B. and {Marchant}, P. and {Schwab}, J. and {Bauer}, E.~B. and {Bildsten}, L. and {Cantiello}, M. and {Dessart}, L. and {Farmer}, R. and {Hu}, H. and {Langer}, N. and {Townsend}, R.~H.~D. and {Townsley}, D.~M. and {Timmes}, F.~X.},
  title = {{Modules for Experiments in Stellar Astrophysics (MESA): Binaries, Pulsations, and Explosions}},
  journal = {\apjs},
  archivePrefix = {arXiv},
  eprint = {1506.03146},
  primaryClass = {astro-ph.SR},
  year = {2015},
  month = {sep},
  volume = {220},
  eid = {15},
  pages = {15},
  doi = {10.1088/0067-0049/220/1/15},
  adsurl = {https://ui.adsabs.harvard.edu/abs/2015ApJS..220...15P},
}

@ARTICLE{Paxton2018,
  author = {{Paxton}, B. and {Schwab}, J. and {Bauer}, E.~B. and {Bildsten}, L. and {Blinnikov}, S. and {Duffell}, P. and {Farmer}, R. and {Goldberg}, J.~A. and {Marchant}, P. and {Sorokina}, E. and {Thoul}, A. and {Townsend}, R.~H.~D. and {Timmes}, F.~X.},
  title = {{Modules for Experiments in Stellar Astrophysics (MESA): Convective Boundaries, Element Diffusion, and Massive Star Explosions}},
  journal = {\apjs},
  archivePrefix = {arXiv},
  eprint = {1710.08424},
  primaryClass = {astro-ph.SR},
  year = {2018},
  month = {feb},
  volume = {234},
  eid = {34},
  pages = {34},
  doi = {10.3847/1538-4365/aaa5a8},
  adsurl = {https://ui.adsabs.harvard.edu/abs/2018ApJS..234...34P},
}

@ARTICLE{Paxton2019,
       author = {{Paxton}, Bill and {Smolec}, R. and {Schwab}, Josiah and {Gautschy}, A. and
         {Bildsten}, Lars and {Cantiello}, Matteo and {Dotter}, Aaron and
         {Farmer}, R. and {Goldberg}, Jared A. and {Jermyn}, Adam S. and
         {Kanbur}, S.~M. and {Marchant}, Pablo and {Thoul}, Anne and
         {Townsend}, Richard H.~D. and {Wolf}, William M. and {Zhang}, Michael and
         {Timmes}, F.~X.},
        title = "{Modules for Experiments in Stellar Astrophysics (MESA): Pulsating Variable Stars, Rotation, Convective Boundaries, and Energy Conservation}",
      journal = {\apjs},
         year = "2019",
        month = "Jul",
       volume = {243},
       number = {1},
          eid = {10},
        pages = {10},
          doi = {10.3847/1538-4365/ab2241},
archivePrefix = {arXiv},
       eprint = {1903.01426},
 primaryClass = {astro-ph.SR},
       adsurl = {https://ui.adsabs.harvard.edu/abs/2019ApJS..243...10P}
}

@ARTICLE{Itoh1996,
  author = {{Itoh}, N. and {Hayashi}, H. and {Nishikawa}, A. and {Kohyama}, Y. },
  title = {{Neutrino Energy Loss in Stellar Interiors. VII. Pair, Photo-, Plasma, Bremsstrahlung, and Recombination Neutrino Processes}},
  journal = {\apjs},
  year = {1996},
  month = {feb},
  volume = {102},
  pages = {411},
  doi = {10.1086/192264},
  adsurl = {https://ui.adsabs.harvard.edu/abs/1996ApJS..102..411I},
}

@article{Chugunov2007,
  author = {{Chugunov}, A.~I. and {Dewitt}, H.~E. and {Yakovlev}, D.~G.},
  title = {{Coulomb tunneling for fusion reactions in dense matter: Path integral MonteCarlo versus mean field}},
  journal = {\prd},
  archivePrefix = {arXiv},
  eprint = {0707.3500},
  year = {2007},
  month = {jul},
  volume = {76},
  number = {2},
  eid = {025028},
  pages = {025028},
  doi = {10.1103/PhysRevD.76.025028},
  adsurl = {https://ui.adsabs.harvard.edu/abs/2007PhRvD..76b5028C},
}

@ARTICLE{Cassisi2007,
   author = {{Cassisi}, S. and {Potekhin}, A.~Y. and {Pietrinferni}, A. and 
	{Catelan}, M. and {Salaris}, M.},
    title = "{Updated Electron-Conduction Opacities: The Impact on Low-Mass Stellar Models}",
  journal = {\apj},
   eprint = {astro-ph/0703011},
     year = 2007,
    month = jun,
   volume = 661,
    pages = {1094-1104},
      doi = {10.1086/516819},
   adsurl = {https://ui.adsabs.harvard.edu/abs/2007ApJ...661.1094C}
}

@ARTICLE{Rogers2002,
   author = {{Rogers}, F.~J. and {Nayfonov}, A.},
    title = "{Updated and Expanded OPAL Equation-of-State Tables: Implications for Helioseismology}",
  journal = {\apj},
     year = 2002,
    month = sep,
   volume = 576,
    pages = {1064-1074},
      doi = {10.1086/341894},
   adsurl = {https://ui.adsabs.harvard.edu/abs/2002ApJ...576.1064R}
}

@ARTICLE{2022MNRAS.514.3058S,
       author = {{Saltas}, Ippocratis D. and {Tognelli}, Emanuele},
        title = "{New calibrated models for the tip of the red giant branch luminosity and a thorough analysis of theoretical uncertainties}",
      journal = {\mnras},
         year = 2022,
        month = aug,
       volume = {514},
       number = {2},
        pages = {3058-3073},
          doi = {10.1093/mnras/stac1546},
archivePrefix = {arXiv},
       eprint = {2203.02499},
 primaryClass = {astro-ph.SR},
       adsurl = {https://ui.adsabs.harvard.edu/abs/2022MNRAS.514.3058S}
}

@ARTICLE{2011arXiv1106.1813C,
       author = {{Chawla}, N.~V. and {Bowyer}, K.~W. and {Hall}, L.~O. and {Kegelmeyer}, W.~P.},
        title = "{SMOTE: Synthetic Minority Over-sampling Technique}",
      journal = {Journal of Artificial Intelligence Research},
         year = 2011,
        month = jun,
          eid = {arXiv:1106.1813},
        pages = {arXiv:1106.1813},
archivePrefix = {arXiv},
       eprint = {1106.1813},
 primaryClass = {cs.AI},
       adsurl = {https://ui.adsabs.harvard.edu/abs/2011arXiv1106.1813C}
}

@ARTICLE{2021ApJ...908L...6R,
       author = {{Riess}, Adam G. and {Casertano}, Stefano and {Yuan}, Wenlong and {Bowers}, J. Bradley and {Macri}, Lucas and {Zinn}, Joel C. and {Scolnic}, Dan},
        title = "{Cosmic Distances Calibrated to 1\% Precision with Gaia EDR3 Parallaxes and Hubble Space Telescope Photometry of 75 Milky Way Cepheids Confirm Tension with {\ensuremath{\Lambda}}CDM}",
      journal = {\apjl},
         year = 2021,
        month = feb,
       volume = {908},
       number = {1},
          eid = {L6},
        pages = {L6},
          doi = {10.3847/2041-8213/abdbaf},
archivePrefix = {arXiv},
       eprint = {2012.08534},
 primaryClass = {astro-ph.CO},
       adsurl = {https://ui.adsabs.harvard.edu/abs/2021ApJ...908L...6R}
}

@ARTICLE{Capozzi2020,
       author = {{Capozzi}, Francesco and {Raffelt}, Georg},
        title = "{Axion and neutrino bounds improved with new calibrations of the tip of the red-giant branch using geometric distance determinations}",
      journal = {\prd},
         year = 2020,
        month = oct,
       volume = {102},
       number = {8},
          eid = {083007},
        pages = {083007},
          doi = {10.1103/PhysRevD.102.083007},
archivePrefix = {arXiv},
       eprint = {2007.03694},
 primaryClass = {astro-ph.SR},
       adsurl = {https://ui.adsabs.harvard.edu/abs/2020PhRvD.102h3007C}
}

@ARTICLE{WortheyLee2011,
       author = {{Worthey}, Guy and {Lee}, Hyun-chul},
        title = "{An Empirical UBV RI JHK Color-Temperature Calibration for Stars}",
      journal = {\apjs},
         year = 2011,
        month = mar,
       volume = {193},
       number = {1},
          eid = {1},
        pages = {1},
          doi = {10.1088/0067-0049/193/1/1},
archivePrefix = {arXiv},
       eprint = {astro-ph/0604590},
 primaryClass = {astro-ph},
       adsurl = {https://ui.adsabs.harvard.edu/abs/2011ApJS..193....1W}
}

@INPROCEEDINGS{1999IAUS..183...48S,
       author = {{Sakai}, S.},
        title = "{The Tip of the Red Giant Branch as a Population II Distance Indicator}",
    booktitle = {Cosmological Parameters and the Evolution of the Universe},
         year = 1999,
       editor = {{Sato}, Katsuhiko},
       volume = {183},
        month = jan,
        pages = {48},
       adsurl = {https://ui.adsabs.harvard.edu/abs/1999IAUS..183...48S}
}

@article{Planck:2018vyg,
    author = "Aghanim, N. and others",
    collaboration = "Planck",
    title = "{Planck 2018 results. VI. Cosmological parameters}",
    eprint = "1807.06209",
    archivePrefix = "arXiv",
    primaryClass = "astro-ph.CO",
    doi = "10.1051/0004-6361/201833910",
    journal = "Astron. Astrophys.",
    volume = "641",
    pages = "A6",
    year = "2020",
    note = "[Erratum: Astron.Astrophys. 652, C4 (2021)]"
}

@ARTICLE{2022arXiv220316683A,
       author = {{Akira Rocha}, Kyle and {Andrews}, Jeff J. and {Berry}, Christopher P.~L. and {Doctor}, Zoheyr and {Marchant}, Pablo and {Kalogera}, Vicky and {Coughlin}, Scott and {Bavera}, Simone S. and {Dotter}, Aaron and {Fragos}, Tassos and {Kovlakas}, Konstantinos and {Misra}, Devina and {Xing}, Zepei and {Zapartas}, Emmanouil},
        title = "{Active Learning for Computationally Efficient Distribution of Binary Evolution Simulations}",
      journal = {arXiv e-prints},
         year = 2022,
        month = mar,
          eid = {arXiv:2203.16683},
        pages = {arXiv:2203.16683},
archivePrefix = {arXiv},
       eprint = {2203.16683},
 primaryClass = {astro-ph.SR},
       adsurl = {https://ui.adsabs.harvard.edu/abs/2022arXiv220316683A}
}

@ARTICLE{1975MSRSL...8..369R,
       author = {{Reimers}, D.},
        title = "{Circumstellar absorption lines and mass loss from red giants.}",
      journal = {Memoires of the Societe Royale des Sciences de Liege},
         year = 1975,
        month = jan,
       volume = {8},
        pages = {369-382},
       adsurl = {https://ui.adsabs.harvard.edu/abs/1975MSRSL...8..369R}
}

@ARTICLE{2010ApJS..189..240C,
       author = {{Cyburt}, Richard H. and {Amthor}, A. Matthew and {Ferguson}, Ryan and
         {Meisel}, Zach and {Smith}, Karl and {Warren}, Scott and {Heger}, Alexand
        er and {Hoffman}, R.~D. and {Rauscher}, Thomas and {Sakharuk}, Alexand
        er and {Schatz}, Hendrik and {Thielemann}, F.~K. and {Wiescher}, Michael},
        title = "{The JINA REACLIB Database: Its Recent Updates and Impact on Type-I X-ray Bursts}",
      journal = {\apjs},
         year = 2010,
        month = jul,
       volume = {189},
       number = {1},
        pages = {240-252},
          doi = {10.1088/0067-0049/189/1/240},
       adsurl = {https://ui.adsabs.harvard.edu/abs/2010ApJS..189..240C}
}

@ARTICLE{2019arXiv191206059L,
       author = {{Liashchynskyi}, Petro and {Liashchynskyi}, Pavlo},
        title = "{Grid Search, Random Search, Genetic Algorithm: A Big Comparison for NAS}",
      journal = {arXiv e-prints},
         year = 2019,
        month = dec,
          eid = {arXiv:1912.06059},
        pages = {arXiv:1912.06059},
archivePrefix = {arXiv},
       eprint = {1912.06059},
 primaryClass = {cs.LG},
       adsurl = {https://ui.adsabs.harvard.edu/abs/2019arXiv191206059L}
}

@ARTICLE{2022arXiv220803651J,
       author = {{Jermyn}, Adam S. and {Bauer}, Evan B. and {Schwab}, Josiah and {Farmer}, R. and {Ball}, Warrick H. and {Bellinger}, Earl P. and {Dotter}, Aaron and {Joyce}, Meridith and {Marchant}, Pablo and {Mombarg}, Joey S.~G. and {Wolf}, William M. and {Wong}, Tin Long Sunny and {Cinquegrana}, Giulia C. and {Farrell}, Eoin and {Smolec}, R. and {Thoul}, Anne and {Cantiello}, Matteo and {Herwig}, Falk and {Toloza}, Odette and {Bildsten}, Lars and {Townsend}, Richard H.~D. and {Timmes}, F.~X.},
        title = "{Modules for Experiments in Stellar Astrophysics (MESA): Time-Dependent Convection, Energy Conservation, Automatic Differentiation, and Infrastructure}",
      journal = {arXiv e-prints},
         year = 2022,
        month = aug,
          eid = {arXiv:2208.03651},
        pages = {arXiv:2208.03651},
archivePrefix = {arXiv},
       eprint = {2208.03651},
 primaryClass = {astro-ph.SR},
       adsurl = {https://ui.adsabs.harvard.edu/abs/2022arXiv220803651J}
}

@ARTICLE{2018A&A...615A..96M,
       author = {{M{\"u}ller}, Oliver and {Rejkuba}, Marina and {Jerjen}, Helmut},
        title = "{Distances from the tip of the red giant branch to the dwarf galaxies dw1335-29 and dw1340-30 in the Centaurus group}",
      journal = {\aap},
         year = 2018,
        month = jul,
       volume = {615},
          eid = {A96},
        pages = {A96},
          doi = {10.1051/0004-6361/201732455},
archivePrefix = {arXiv},
       eprint = {1803.02406},
 primaryClass = {astro-ph.GA},
       adsurl = {https://ui.adsabs.harvard.edu/abs/2018A&A...615A..96M}
}

@article{Dolan:2021rya,
    author = "Dolan, Matthew J. and Hiskens, Frederick J. and Volkas, Raymond R.",
    title = "{Constraining axion-like particles using the white dwarf initial-final mass relation}",
    eprint = "2102.00379",
    archivePrefix = "arXiv",
    primaryClass = "hep-ph",
    doi = "10.1088/1475-7516/2021/09/010",
    journal = "JCAP",
    volume = "09",
    pages = "010",
    year = "2021"
}

@article{Corsico:2019nmr,
    author = "C\'orsico, Alejandro H. and Althaus, Leandro G. and Miller Bertolami, Marcelo M. and Kepler, S. O.",
    title = "{Pulsating white dwarfs: new insights}",
    eprint = "1907.00115",
    archivePrefix = "arXiv",
    primaryClass = "astro-ph.SR",
    doi = "10.1007/s00159-019-0118-4",
    journal = "Astron. Astrophys. Rev.",
    volume = "27",
    number = "1",
    pages = "7",
    year = "2019"
}

@article{MillerBertolami:2014rka,
    author = "Miller Bertolami, Marcelo M. and Melendez, Brenda E. and Althaus, Leandro G. and Isern, Jordi",
    title = "{Revisiting the axion bounds from the Galactic white dwarf luminosity function}",
    eprint = "1406.7712",
    archivePrefix = "arXiv",
    primaryClass = "hep-ph",
    doi = "10.1088/1475-7516/2014/10/069",
    journal = "JCAP",
    volume = "10",
    pages = "069",
    year = "2014"
}

@article{Ayala:2014pea,
    author = "Ayala, Adrian and Dom\'\i{}nguez, Inma and Giannotti, Maurizio and Mirizzi, Alessandro and Straniero, Oscar",
    title = "{Revisiting the bound on axion-photon coupling from Globular Clusters}",
    eprint = "1406.6053",
    archivePrefix = "arXiv",
    primaryClass = "astro-ph.SR",
    doi = "10.1103/PhysRevLett.113.191302",
    journal = "Phys. Rev. Lett.",
    volume = "113",
    number = "19",
    pages = "191302",
    year = "2014"
}

@article{Carenza:2020zil,
    author = {Carenza, Pierluca and Straniero, Oscar and D\"obrich, Babette and Giannotti, Maurizio and Lucente, Giuseppe and Mirizzi, Alessandro},
    title = "{Constraints on the coupling with photons of heavy axion-like-particles from Globular Clusters}",
    eprint = "2004.08399",
    archivePrefix = "arXiv",
    primaryClass = "hep-ph",
    doi = "10.1016/j.physletb.2020.135709",
    journal = "Phys. Lett. B",
    volume = "809",
    pages = "135709",
    year = "2020"
}

@article{Friedland:2012hj,
    author = "Friedland, Alexander and Giannotti, Maurizio and Wise, Michael",
    title = "{Constraining the Axion-Photon Coupling with Massive Stars}",
    eprint = "1210.1271",
    archivePrefix = "arXiv",
    primaryClass = "hep-ph",
    reportNumber = "LA-UR-12-25074",
    doi = "10.1103/PhysRevLett.110.061101",
    journal = "Phys. Rev. Lett.",
    volume = "110",
    number = "6",
    pages = "061101",
    year = "2013"
}

@article{Sakstein:2022tby,
    author = "Sakstein, Jeremy and Croon, Djuna and McDermott, Samuel D.",
    title = "{Axion instability supernovae}",
    eprint = "2203.06160",
    archivePrefix = "arXiv",
    primaryClass = "hep-ph",
    reportNumber = "IPPP/22/10, FERMILAB-PUB-22-118-T",
    doi = "10.1103/PhysRevD.105.095038",
    journal = "Phys. Rev. D",
    volume = "105",
    number = "9",
    pages = "095038",
    year = "2022"
}

@article{Baxter:2021swn,
    author = "Baxter, Eric J. and Croon, Djuna and McDermott, Samuel D. and Sakstein, Jeremy",
    title = "{Find the Gap: Black Hole Population Analysis with an Astrophysically Motivated Mass Function}",
    eprint = "2104.02685",
    archivePrefix = "arXiv",
    primaryClass = "astro-ph.CO",
    reportNumber = "FERMILAB-PUB-21-148-T, IPPP/20/88",
    doi = "10.3847/2041-8213/ac11fc",
    journal = "Astrophys. J. Lett.",
    volume = "916",
    number = "2",
    pages = "L16",
    year = "2021"
}

@article{Sakstein:2020axg,
    author = "Sakstein, Jeremy and Croon, Djuna and McDermott, Samuel D. and Straight, Maria C. and Baxter, Eric J.",
    title = "{Beyond the Standard Model Explanations of GW190521}",
    eprint = "2009.01213",
    archivePrefix = "arXiv",
    primaryClass = "gr-qc",
    reportNumber = "FERMILAB-PUB-20-461-T",
    doi = "10.1103/PhysRevLett.125.261105",
    journal = "Phys. Rev. Lett.",
    volume = "125",
    number = "26",
    pages = "261105",
    year = "2020"
}

@article{Straight:2020zke,
    author = "Straight, Maria C. and Sakstein, Jeremy and Baxter, Eric J.",
    title = "{Modified Gravity and the Black Hole Mass Gap}",
    eprint = "2009.10716",
    archivePrefix = "arXiv",
    primaryClass = "gr-qc",
    doi = "10.1103/PhysRevD.102.124018",
    journal = "Phys. Rev. D",
    volume = "102",
    pages = "124018",
    year = "2020"
}

@ARTICLE{2022arXiv220801110C,
       author = {{Croon}, Djuna and {Sakstein}, Jeremy},
        title = "{Light Axion Emission and the Formation of Merging Binary Black Holes}",
      journal = {arXiv e-prints},
         year = 2022,
        month = aug,
          eid = {arXiv:2208.01110},
        pages = {arXiv:2208.01110},
archivePrefix = {arXiv},
       eprint = {2208.01110},
 primaryClass = {astro-ph.HE},
       adsurl = {https://ui.adsabs.harvard.edu/abs/2022arXiv220801110C}
}

@ARTICLE{dennis2025machinelearningtipred,
       author = {{Dennis}, Mitchell and {Sakstein}, Jeremy},
        title = "{Machine Learning the Tip of the Red Giant Branch}",
      journal = {arXiv e-prints},
         year = 2025,
        month = sep,
          eid = {arXiv:2303.12069},
        pages = {arXiv:2303.12069},
          doi = {10.48550/arXiv.2303.12069},
archivePrefix = {arXiv},
       eprint = {2303.12069},
 primaryClass = {astro-ph.GA},
       adsurl = {https://ui.adsabs.harvard.edu/abs/2023arXiv230312069D}
}

@dataset{dennis_mitchell_t_2023_7896061,
  author       = {Dennis, Mitchell T. and
                  Sakstein, Jeremy},
  title        = {{Tip of the Red Giant Branch Bounds on the Axion- 
                   Electron Coupling Revisited}},
  month        = may,
  year         = 2023,
  publisher    = {Zenodo},
  version      = 1,
  doi          = {10.5281/zenodo.7896061},
  url          = {https://doi.org/10.5281/zenodo.7896061}
}

@ARTICLE{2023arXiv230713050F,
       author = {{Franz}, Noah and {Dennis}, Mitchell and {Sakstein}, Jeremy},
        title = "{Tip of the Red Giant Branch Bounds on the Neutrino Magnetic Dipole Moment Revisited}",
      journal = {arXiv e-prints},
         year = 2023,
        month = jul,
          eid = {arXiv:2307.13050},
        pages = {arXiv:2307.13050},
          doi = {10.48550/arXiv.2307.13050},
archivePrefix = {arXiv},
       eprint = {2307.13050},
 primaryClass = {hep-ph},
       adsurl = {https://ui.adsabs.harvard.edu/abs/2023arXiv230713050F}
}

@article{Altherr:1993zd,
    author = "Altherr, T. and Petitgirard, E. and del Rio Gaztelurrutia, T.",
    title = "{Axion emission from red giants and white dwarfs}",
    eprint = "hep-ph/9310304",
    archivePrefix = "arXiv",
    reportNumber = "CERN-TH-7044-93, ENSLAPP-A-441-93",
    doi = "10.1016/0927-6505(94)90040-X",
    journal = "Astropart. Phys.",
    volume = "2",
    pages = "175--186",
    year = "1994"
}

@ARTICLE{1996ApJ...464..943I,
       author = {{Iglesias}, Carlos A. and {Rogers}, Forrest J.},
        title = "{Updated Opal Opacities}",
      journal = {\apj},
         year = 1996,
        month = jun,
       volume = {464},
        pages = {943},
          doi = {10.1086/177381},
       adsurl = {https://ui.adsabs.harvard.edu/abs/1996ApJ...464..943I}
}

@BOOK{1968pss..book.....C,
       author = {{Cox}, J.~P. and {Giuli}, R.~T.},
        title = "{Principles of stellar structure}",
         year = 1968,
       adsurl = {https://ui.adsabs.harvard.edu/abs/1968pss..book.....C}
}

@article{Croon:2023trk,
    author = "Croon, Djuna and Sakstein, Jeremy",
    title = "{Dark matter annihilation and pair-instability supernovae}",
    eprint = "2310.20044",
    archivePrefix = "arXiv",
    primaryClass = "astro-ph.HE",
    doi = "10.1103/PhysRevD.109.103021",
    journal = "Phys. Rev. D",
    volume = "109",
    number = "10",
    pages = "103021",
    year = "2024"
}

@article{Troitsky:2024keu,
    author = "Troitsky, S. V.",
    title = "{Stellar evolution and axion-like particles: new constraints and hints from globular clusters in the GAIA DR3 data}",
    eprint = "2410.02266",
    archivePrefix = "arXiv",
    primaryClass = "hep-ph",
    reportNumber = "INR-TH-2024-018",
    month = "10",
    year = "2024"
}

@article{Frieman:1987ui,
    author = "Frieman, Joshua A. and Dimopoulos, Savas and Turner, Michael S.",
    title = "{Axions and Stars}",
    reportNumber = "SLAC-PUB-4172, FERMILAB-PUB-87-254-A",
    doi = "10.1103/PhysRevD.36.2201",
    journal = "Phys. Rev. D",
    volume = "36",
    pages = "2201",
    year = "1987"
}
